\documentclass[superscriptaddress, showpacs, twocolumn, secnumarabic, amssymb, nobibnotes, aps, prb, longbibliography]{revtex4-1}

\usepackage{graphicx}
\usepackage{dcolumn}
\usepackage{subfigure}
\usepackage{bm}
\usepackage{float}
\usepackage{amssymb}
\usepackage{amsmath}
\usepackage{comment}
\usepackage[countmax]{subfloat}

\makeatletter
\newcommand{\ssymbol}[1]{^{\@fnsymbol{#1}}}
\makeatother

\begin{document}

\title{Isoelectronic perturbations to \textit{f}-\textit{d}-electron hybridization and the enhancement of hidden order in URu\textsubscript{2}Si\textsubscript{2}}

\author{C. T. Wolowiec} 
\affiliation{Department of Physics, University of California, San Diego, La Jolla, California 92093, USA}
\affiliation{Center for Advanced Nanoscience, University of California, San Diego, La Jolla, California 92093, USA}
\author{N. Kanchanavatee}
\affiliation{Department of Physics, Chulalongkorn University, Pathumwan 10330, Thailand}
\author{K. Huang}
\affiliation{Lawrence Livermore National Laboratory, Livermore, California 94550, USA}
\author{S. Ran}
\affiliation{Department of Physics, Washington University in St. Louis, St. Louis, Missouri 63130, USA}
\author{A. J. Breindel}
\affiliation{Department of Physics, University of California, San Diego, La Jolla, California 92093, USA}
\affiliation{Center for Advanced Nanoscience, University of California, San Diego, La Jolla, California 92093, USA}
\author{N. Pouse}
\affiliation{Department of Physics, University of California, San Diego, La Jolla, California 92093, USA}
\author{Kalyan Sasmal}
\affiliation{Department of Physics, University of California, San Diego, La Jolla, California 92093, USA}
\affiliation{Center for Advanced Nanoscience, University of California, San Diego, La Jolla, California 92093, USA}
\author{R. E. Baumbach}
\affiliation{National High Magnetic Field Laboratory, Florida State University, Tallahassee, Florida 32310, USA}
\affiliation{Department of Physics, Florida State University, Tallahassee, Florida 32306, USA}
\author{G. Chappell}
\affiliation{National High Magnetic Field Laboratory, Florida State University, Tallahassee, Florida 32310, USA}
\affiliation{Department of Physics, Florida State University, Tallahassee, Florida 32306, USA}
\author{Peter S. Riseborough}
\affiliation{Physics Department, Temple University, Philadelphia, PA 19122, USA}
\author{M. B. Maple}
\email[Corresponding Author: ]{mbmaple@ucsd.edu}
\affiliation{Department of Physics, University of California, San Diego, La Jolla, California 92093, USA}
\affiliation{Center for Advanced Nanoscience, University of California, San Diego, La Jolla, California 92093, USA}
\date{\today}

\begin{abstract}
Electrical resistivity measurements were performed on single crystals of URu\textsubscript{2-\textit{x}}Os\textsubscript{\textit{x}}Si\textsubscript{2} up to \textit{x} = 0.28 under hydrostatic pressure up to \textit{P} = 2 GPa. As the Os concentration, \textit{x}, is increased, (1) the lattice expands, creating an effective \textit{negative} chemical pressure  \textit{P}\textsubscript{ch}(\textit{x}), (2) the hidden order (HO) phase is enhanced and the system is driven toward a large-moment antiferromagnetic (LMAFM) phase, and (3) less external pressure \textit{P}\textsubscript{c} is required to induce the HO$\rightarrow$LMAFM phase transition. We compare the \textit{T}(\textit{x}, \textit{P}) phase behavior reported here for the URu\textsubscript{2-\textit{x}}Os\textsubscript{\textit{x}}Si\textsubscript{2} system with previous reports of enhanced HO in URu\textsubscript{2}Si\textsubscript{2} upon tuning with \textit{P}, or similarly in URu\textsubscript{2-\textit{x}}Fe\textsubscript{\textit{x}}Si\textsubscript{2} upon tuning with \textit{positive} \textit{P}\textsubscript{ch}(\textit{x}). It is noted that pressure, Fe substitution, and Os substitution are the only known perturbations that enhance the HO phase and induce the first order transition to the LMAFM phase in URu\textsubscript{2}Si\textsubscript{2}. We present a scenario in which the application of pressure or the isoelectronic substitution of Fe and Os ions for Ru results in an increase in the hybridization of the U-5\textit{f}- and transition metal \textit{d}-electron states which leads to electronic instability in the paramagnetic phase and a concurrent stability of HO (and  LMAFM) in URu\textsubscript{2}Si\textsubscript{2}. Calculations in the tight binding approximation are included to determine the strength of hybridization between the U-5\textit{f} electrons and each of the isoelectronic transition metal \textit{d}-electron states of Fe, Ru, and Os.
\end{abstract}

\pacs{71.27.+a, 72.10.Di, 74.62.Dh, 74.62.Fj} 

\maketitle
\section{introduction}
The heavy-fermion superconducting compound URu$_2$Si$_2$ is known for its second-order phase transition into the so-called ``hidden order'' (HO) phase at a transition temperature $T_0$ $\approx$ 17.5 K.
Extensive investigation of the phase space in proximity to the HO phase transition has provided a detailed picture of the electronic and magnetic structure of this unique phase.\cite{Palstra_1985, Maple_1986, Schlabitz_1986, Broholm_1987, Schoenes_1987,McElfresh_1987,Dawson_1989, Mason_1991, Broholm_1991, Santini_1994, Buyers_1994, Escudero_1994, Amitsuka_1999,  Chandra_2002, Bourdarot_2003, Wiebe_2004, Behnia_2005, Jeffries_2007, Wiebe_2007, Elgazzar_2009, Janik_2009, Santander_2009, Haule_2010, Yoshida_2010, Schmidt_2010, Aynajian_2010, Oppeneer_2010, Oppeneer_2011, Kawasaki_2011, Dakovski_2011, Dubi_2011, Haraldsen_2011, Pepin_2011, Riseborough_2012, Mydosh_2011, Meng_2013, Boariu_2013, Bareille_2014, Mydosh_2014, Butch_2015, Ran_2016, Kung_2016} However, more than three decades after the initial characterization of URu$_2$Si$_2$,\cite{Palstra_1985, Maple_1986, Schlabitz_1986} the order parameter for the HO phase is still unidentified.\\
\indent Most perturbations to the URu$_{2}$Si$_2$ compound have the effect of suppressing HO. The application of an external magnetic field ($H$) suppresses the HO phase\cite{Ran_2016, Ran_2017} and many of the chemical substitutions ($x$) at the U, Ru, or Si sites that have been explored, significantly reduce $T_0$, even at modest levels of substituent concentration.\cite{Dalichaouch_1989, Dalichaouch_1990, Dalichaouch_1990b, Butch_2010b, Yokoyama_2004, delaTorre_1992, Park_1994, Zwirner_1997, Gallagher_2015} 
At present, only three perturbations are known to consistently \textit{enhance} the HO phase in URu$_{2}$Si$_2$: (1) external pressure $P$, (2) isoelectronic substitution of Fe ions for Ru, and (3) isoelectronic substitution of Os ions for Ru. Upon applying pressure $P$, the HO phase in pure URu$_2$Si$_2$ is enhanced\cite{McElfresh_1987} and the system is driven toward a large moment antiferromagnetic (LMAFM) phase.\cite{Butch_2010a} The HO$\rightarrow$LMAFM phase transition is identified indirectly by a characteristic ``kink'' at a critical pressure $P_c$ $\approx$ 1.5 GPa in the $T_0$($P$) phase boundary,\cite{Jeffries_2007, Jeffries_2008, Butch_2010a} and also directly by neutron diffraction experiments, which reveal an increase in the magnetic moment from $\mu$ $\sim$ ($0.03 \pm 0.02)$$\mu_B$/U in the HO phase to $\mu$ $\sim$ $0.4$ $\mu_B$/U in the LMAFM phase.\cite{Amitsuka_1999, Bourdarot_2004, Amitsuka_2007}\\ 
\indent Recent reports indicate that the isoelectronic substitution of Fe ions for Ru in URu$_2$Si$_2$ replicates the $T_0(P)$ behavior in URu$_2$Si$_2$.\cite{Kanchanavatee_2011, Janoschek_2015, Wolowiec_2016} An increase in $x$ in URu$_{2-x}$Fe$_x$Si$_2$ enhances the HO phase and drives the system toward the HO$\rightarrow$LMAFM phase transition at a critical Fe concentration $x_c$ $\approx$ 0.15.\cite{Janoschek_2015, Wilson_2016} The decrease in the volume of the unit cell due to substitution of smaller Fe ions for Ru may be interpreted as a chemical pressure, $P_{ch}$, where the Fe concentration $x$ can be converted to $P_{ch}$($x$).\cite{Kanchanavatee_2011, Wolowiec_2016} In addition, the induced HO$\rightarrow$LMAFM phase transition in URu$_{2-x}$Fe$_x$Si$_2$ occurs at combinations of $x$ and $P$ that consistently obey the additive relationship: $P_{ch}(x)$ + $P_c$ $\approx$ 1.5 GPa. \cite{Kanchanavatee_2011, Wolowiec_2016} These results have led to the suggestion that $P_{ch}$ is equivalent to $P$ in affecting the HO and LMAFM phases.\cite{Janoschek_2015, Wolowiec_2016}\\
\indent Reports of the isoelectronic substitution of larger Os ions for Ru have shown that an increase in $x$ in URu$_{2-x}$Os$_x$Si$_2$: (1) expands the volume of the unit cell, thus creating an effective $negative$ chemical pressure ($P_{ch}$ $\leq$ 0), (2) enhances the HO phase, and (3) drives the system toward a similar HO$\rightarrow$LMAFM phase transition at a critical Os concentration of $x_c$ $\approx$ 0.065. \cite{Kanchanavatee_2014, Hall_2015, Wilson_2016} These results are contrary to the expectation that a $negative$ $P_{ch}$ would lead to a suppression of HO and complicates the view of ``chemical pressure'' as a mechanism affecting the evolution of phases in URu$_{2}$Si$_2$.\\
\indent In this paper, we report on the $T$($x$, $P$) phase behavior for the URu$_{2-x}$Os$_x$Si$_2$ system based on $\rho(T)$ measurements of single crystals of URu$_{2-x}$Os$_x$Si$_2$ as a function of Os concentration $x$, and applied pressure $P$. The $T$($x$, $P$) phase behavior observed here for the URu$_{2-x}$Os$_{x}$Si$_{2}$ system\cite{Kanchanavatee_2011, Janoschek_2015, Wolowiec_2016} is compared to that of the URu$_{2-x}$Fe$_x$Si$_2$ system and also with the $T(P)$ behavior in pure URu$_{2}$Si$_2$. As an explanation for the enhancement of HO toward the HO$\rightarrow$LMAFM phase transition, we suggest a scenario in which each of the perturbations of Os substitution, Fe substitution, and pressure $P$ favors delocalization of the 5$f$ electrons and increases the hybridization of the uranium 5$f$- and transition metal (Fe, Ru, Os) $d$-electron states. In order to avoid an ad-hoc explanation of the effect of increasing the Os concentration $x$ in URu$_{2-x}$Os$_x$Si$_2$, compared to the effects of pressure $P$ and Fe substitution, we explain how pressure $P$, Fe substitution, and Os substitution are three perturbative routes to enhancement of the U-5$f$- and $d$-electron hybridization. The importance of the 5$f$- and $d$-electron hybridization to the emergence of HO/LMAFM is presented in the context of the Fermi surface instability that leads to a reconstruction and partial gapping of the Fermi-surface during the transition from the paramagnetic (PM) phase to the HO and LMAFM phases.\cite{Maple_1986, McElfresh_1987, Denlinger_2001, Santander_2009, Elgazzar_2009, Yoshida_2010, Schmidt_2010, Aynajian_2010, Boariu_2013, Bareille_2014}\\
\indent In an effort to further understand the effect of isoelectronic substitution on the 5$f$- and $d$-electron hybridization, calculations in the tight-binding approximation were made for compounds from the series U$M$$_{2}$Si$_2$  ($M$ = Fe, Ru, and Os). The calculations indicate that the degree of hybridization is largely dependent on the magnitude of the difference between the binding energy of the localized U-5$f$ electrons and that of the transition metal $d$ electrons. 

\section{Experimental Details}
\label{Experimental Details}
\indent The experimental design and procedure, including synthesis of single crystals, crystallographic measurements, and measurements of electrical resistivity under applied pressure are similar to those in the investigation of the URu$_{2-x}$Fe$_x$Si$_2$ system as described in Ref.~\onlinecite{Wolowiec_2016}. Single crystals of URu$_{2-x}$Os$_x$Si$_2$ at nominal concentrations of $x_{nom}$ = 0, 0.025, 0.05, 0.10, 0.13, 0.16, and 0.20 were grown according to the Czochralski method in a tetra-Arc furnace. The quality of the single crystal samples were determined by Laue X-ray diffraction patterns together with X-ray powder diffraction (XRD)  measurements. The XRD patterns were fitted according to the Rietveld refinement technique using the {\it GSAS-II} software package. \cite{Toby_2013} Elemental analysis of single crystal samples of URu$_{2-x}$Os$_x$Si$_2$ at nominal concentrations of $x_{nom}$ = 0.025, 0.05, 0.10, 0.13, and 0.20 was performed using energy-dispersive X-ray spectroscopy (EDX). Based on the EDX measurements, the actual osmium concentrations $x_{act}$ in these samples were determined to be  $x_{act}$ = 0.07, 0.08, 0.15, 0.18, and 0.28, respectively. In this report, the Os concentration $x$ in these single crystal URu$_{2-x}$Os$_x$Si$_2$ samples is taken as $x_{act}$ as determined from the EDX measurements, unless otherwise stated. It is noted that the single crystal sample with nominal Os concentration $x_{nom}$ = 0.16 was not available for EDX measurement and thus $x$ = $x_{nom}$ = 0.16 in this case. (See Sec.~\ref{appendix} for details on sample quality and the error in Os concentration.)\\
\begin{figure}[b]
\includegraphics[scale = 0.38, trim= 3cm 1cm 1cm 2cm, clip=true]{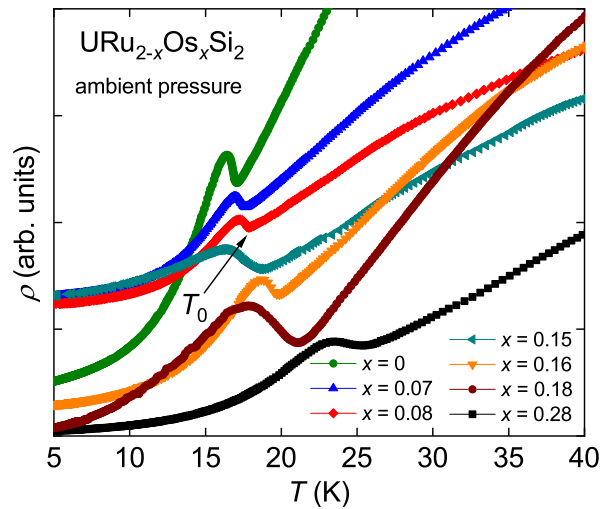}
\vspace{- 0.5cm}
\caption{\label{ambient resistivity}  (Color online) Electrical resistivity $\rho(T)$ in the vicinity of the HO/LMAFM transition for the URu$_{2-x}$Os$_{x}$Si$_{2}$ system at ambient pressure for $x$ = 0, 0.07, 0.08, 0.15, 0.16, 0.18, and 0.28. The transition temperature $T_0$ is indicated by the black arrow. The $\rho(T)$ curves have been shifted vertically for clarity.}
\vspace{0.3cm}
\end{figure}
\indent Annealed Pt wire leads were affixed with silver epoxy to gold-sputtered contact surfaces on each sample in a standard four-wire configuration. Electrical resistivity $\rho$($T$) measurements were performed on single crystals of URu$_{2-x}$Os$_x$Si$_2$ under applied pressure up to $P$ = 2 GPa for Os concentrations $x$ = 0, 0.07, 0.08, 0.15, 0.16, 0.18 and 0.28. A 1:1 mixture by volume of $n$-pentane and isoamyl alcohol was used to provide a quasi-hydrostatic pressure transmitting medium and the pressure was locked in with the use of a beryllium copper clamped piston-cylinder pressure cell. The pressure dependence of the superconducting transition temperature of high purity Sn was used as a manometer. Measurements of $\rho(T)$ were performed upon warming from $\sim$ 1 to 300 K in a pumped $^4$He dewar and the temperature was determined from the four-wire electrical resistivity of a calibrated Cernox sensor.
\begin{figure}[b]
\includegraphics[scale = 0.389, trim= 2.5cm 1.5cm 2.7cm 2.3cm, clip=true]{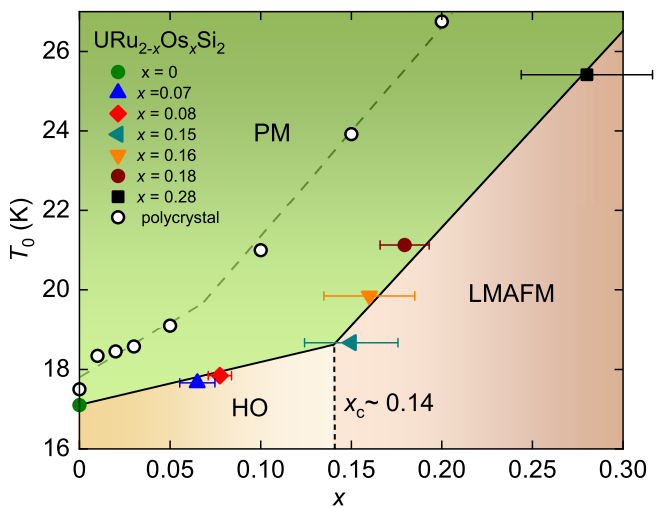}
\vspace{- 0.5cm}
\caption{\label{phase_diagram_x}(Color online) $T_0$ vs $x$ phase diagram for URu$_{2-x}$Os$_x$Si$_2$ up to $x$ = 0.28. The solid black and green dashed lines representing the $T_0(x)$ phase boundaries are linear fits to the values of $T_0$ for the single crystal and polycrystalline samples, respectively. The values of $T_0$ were determined from $\rho(T)$ data as shown in Fig.~\ref{ambient resistivity} (see text) and the values of  $T_0$ for the polycrystalline samples (white circles) were similarly determined as reported in Ref.~\onlinecite{Kanchanavatee_2014}. The vertical dashed line locates the critical Os concentration $x_c$ $\approx$ 0.14 at the HO$\rightarrow$LMAFM phase transition. Error bars for $x$ represent standard deviations in the data from EDX measurements (see Sec.~\ref{appendix}).}
\vspace{- 0.1cm}
\end{figure}
\vspace{- 0.5cm}
\section{Results}
\indent Figure~\ref{ambient resistivity} displays the temperature dependence of the ambient pressure electrical resistivity $\rho(T)$ in the vicinity of the transition temperature $T_0$ for the URu$_{2-x}$Os$_x$Si$_2$ system. The transition from the paramagnetic (PM) phase to the HO phase (or LMAFM phase at higher values of $x$) is defined to be at the location of the minimum in $\rho(T)$, which occurs prior to the upturn in $\rho(T)$ upon cooling, as indicated by the black arrow. It is clear that the feature in $\rho(T)$ shifts to higher temperature as $x$ is increased.\\
\begin{figure}[t]
\includegraphics[scale = 0.3, trim= 1.8cm 0.7cm 1cm 2cm, clip=true]{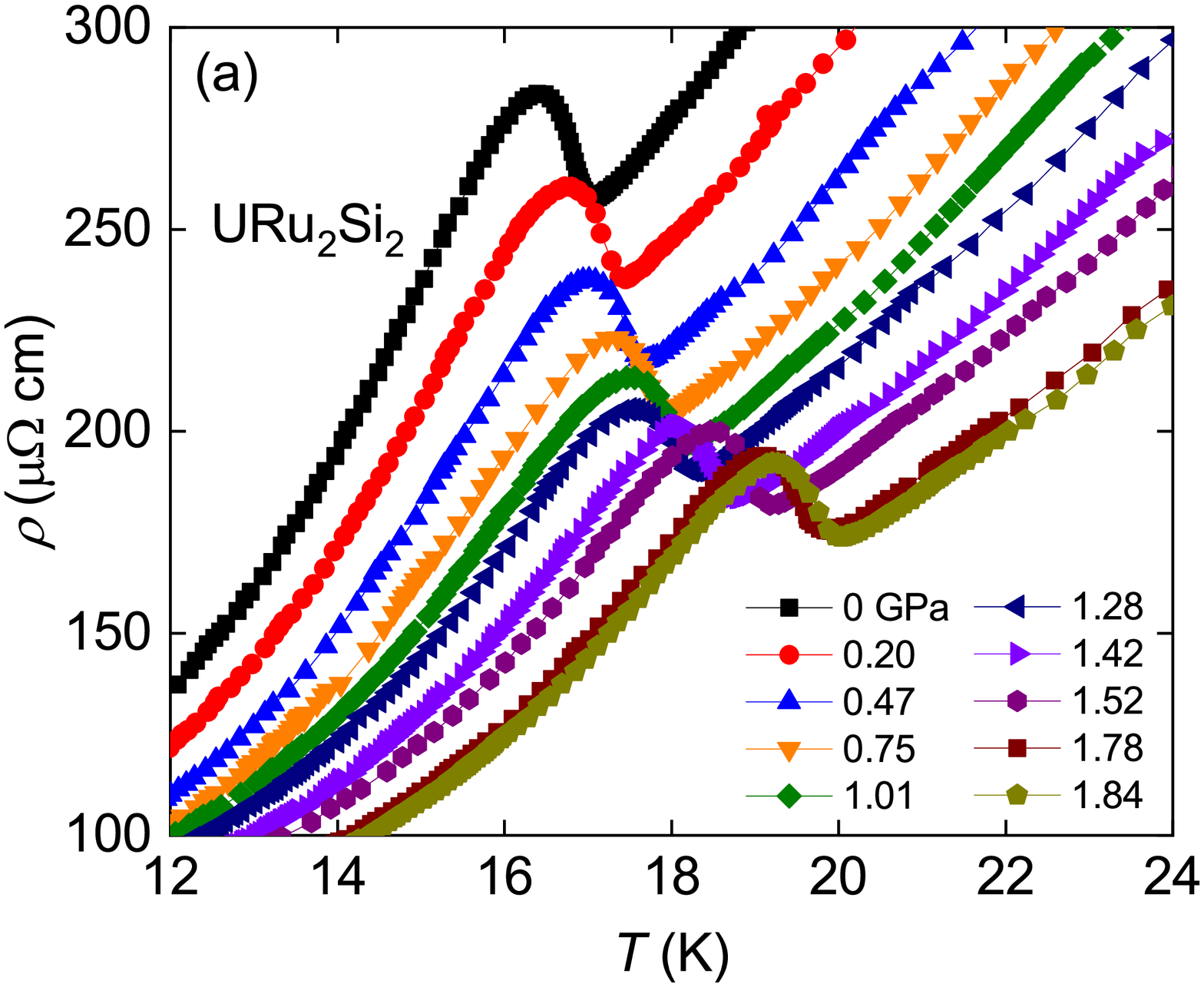}
\includegraphics[scale = 0.3, trim= 1.8cm 0.8cm 1cm 2cm, clip=true]{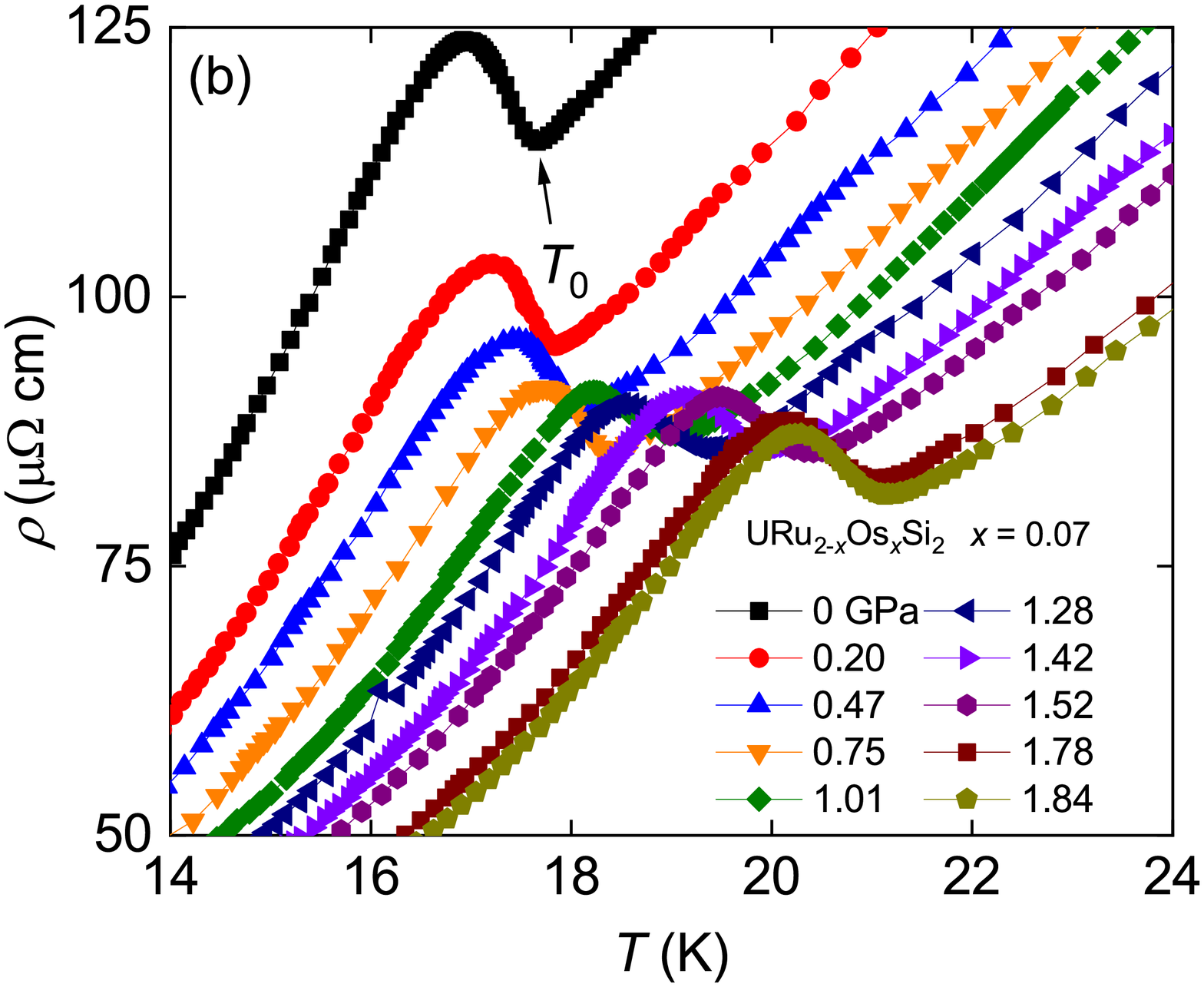}
\includegraphics[scale = 0.293, trim= 1.6cm 1.0cm 0.9cm 1.7cm, clip=true]{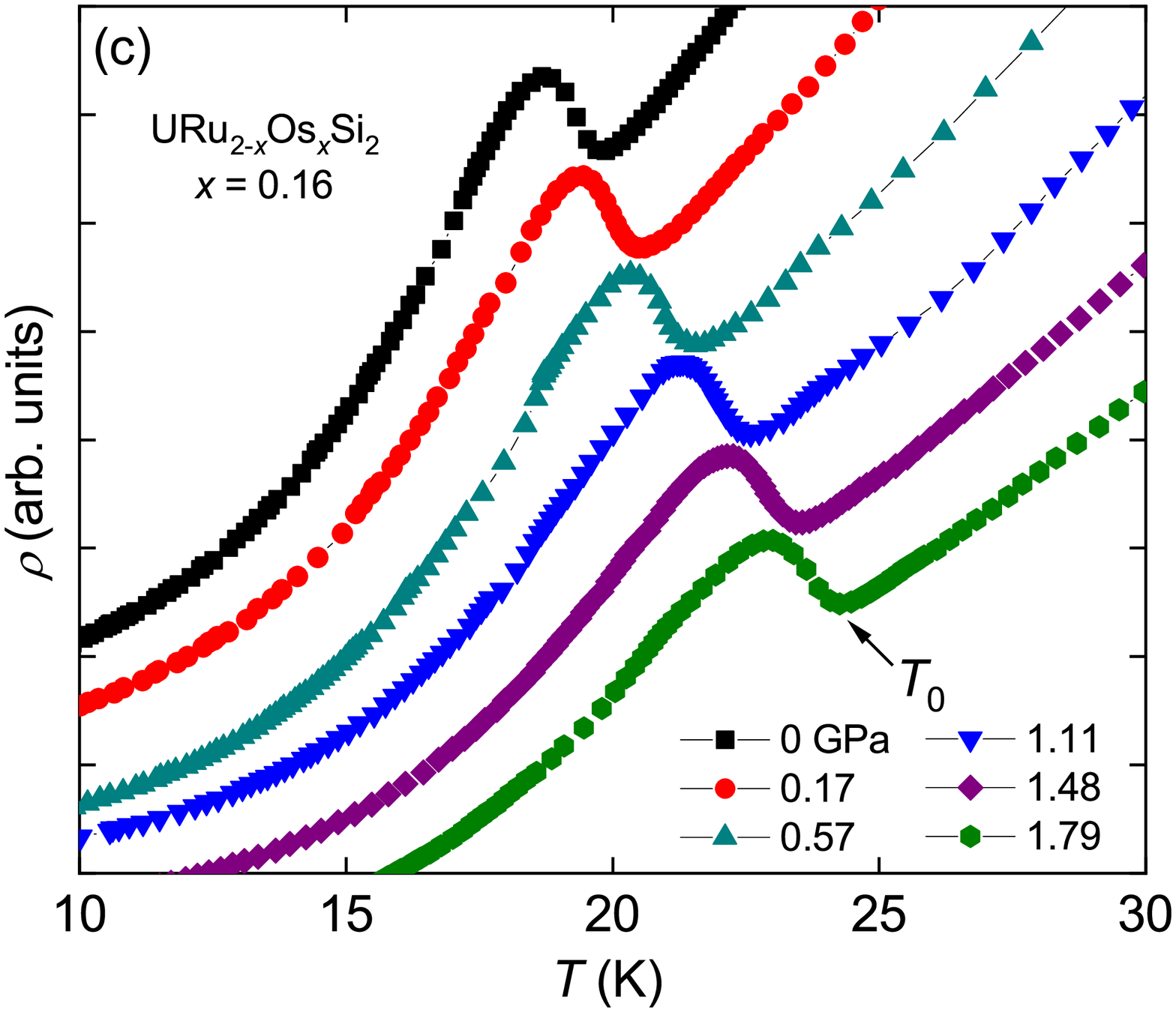}
\caption{\label{pressure resistivity}  (Color online) Electrical resistivity $\rho(T)$ in the vicinity of the HO/LMAFM transition for the URu$_{2-x}$Os$_{x}$Si$_{2}$ system as a function of pressure $P$: (a) $\rho(T)$ for pure URu$_{2}$Si$_2$ as a function of pressure up to $P$ = 1.9 GPa; (b)  $\rho(T)$ for the $x$ = 0.07 sample as a function of pressure up to $P$ = 1.9 GPa; and (c) $\rho(T)$ for the $x$ = 0.16 sample as a function of pressure up to 1.8 GPa. The $\rho(T)$ data for $x$ = 0.16 have been shifted vertically for clarity.}
\end{figure}
\indent The values of $T_0$, as determined from the $\rho(T)$ data shown in Fig.~\ref{ambient resistivity} for single crystal samples of URu$_{2-x}$Os$_{x}$Si$_{2}$ at $x$ = 0, 0.07, 0.08, 0.15, 0.16, 0.18, and 0.28, were used to construct the $T$--$x$ phase diagram displayed in Fig.~\ref{phase_diagram_x}. The solid black lines that outline the $T_0(x)$ phase boundary between the PM phase and the HO (or LMAFM) phase are linear fits to the $T_0(x)$ data. The solid black line of smaller slope outlining the $T_0(x)$ phase boundary between the PM phase and the HO phase is a linear fit to the $T_0(x)$ data for samples with low Os concentrations up to $x$ = 0.15. The solid black line of larger slope outlining the $T_0(x)$ phase boundary between the PM phase and the LMAFM phase is a linear fit to the $T_0(x)$ data for the single crystal samples with Os concentrations from $x$ = 0.15 to 0.28. The intersection of the two lines forms a ``kink'' in the $T_0(x)$ phase boundary and is taken to be the location of the HO$\rightarrow$LMAFM transition at a critical Os concentration of $x_c$ $\approx$ 0.14.\\
\indent Similar linear fits (dashed green lines) to the values of $T_0$ (white circles) taken from Ref.~\onlinecite{Kanchanavatee_2014} for polycrystalline samples of URu$_{2-x}$Os$_{x}$Si$_{2}$ suggest a critical concentration of $x_c$ $\approx$ 0.07. 
Given the differences inherent to synthesis of polycrystals compared to single crystals, the discrepancy between the values of the critical concentration for the polycrystalline and single crystal samples is not clearly understood. However, the two $T_0(x)$ phase boundaries for polycrystalline and single crystal samples of URu$_{2-x}$Os$_x$Si$_2$ are qualitatively similar with the $T_0(x)$ phase boundary for single crystals being shifted toward higher Os concentration $x$.\\
\begin{figure}[t] 
 \includegraphics[width=1.0\linewidth, trim= 2cm 1cm 2cm 2cm, clip=true]{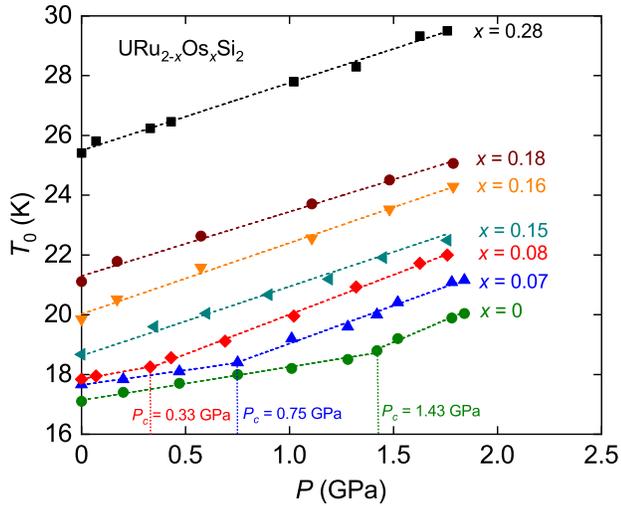}
 \vspace{- 0.4cm}
\caption{\label{phase} (Color online) A ``composite'' $T_0$ vs $P$ phase diagram for URu$_{2-x}$Os$_{x}$Si$_{2}$. The dashed lines representing the $T_0(P)$ phase boundary are linear fits to the $T_0(P)$ data. The values of the critical pressures $P_{c}$ = 1.43, 0.75, 0.33 for $x$ = 0, 0.07, and 0.08, respectively, mark the pressure-induced HO$\rightarrow$LMAFM phase transition and are defined by the location of the ``kinks'' in the $T_0(P)$ phase boundaries (or intersections of the linear fits) (see text).} 
\vspace{-.37 cm}
\end{figure}  
\indent Figure~\ref{pressure resistivity} displays the temperature dependence of the electrical resistivity $\rho(T)$ near $T_0$ for the URu$_{2-x}$Os$_x$Si$_2$ system under applied pressure $P$. Figure~\ref{pressure resistivity}(a) displays $\rho(T)$ for pure URu$_{2}$Si$_2$ as a function of pressure up to $P$ $\approx$ 1.9 GPa. As pressure is increased, the feature in $\rho(T)$ shifts to higher temperature similar to what is observed with an increase in $x$. Furthermore, the feature in $\rho(T)$ appears to migrate more quickly with pressure above some critical pressure near 1.4 GPa. Figures~\ref{pressure resistivity}(b) and (c) display $\rho(T)$ in the vicinity of $T_0$ as a function of applied pressure for samples at $x$ = 0.07 and 0.16, respectively. The sample with $x$ = 0.07 (Fig.~\ref{pressure resistivity}(b)) is at an Os concentration well below the critical concentration $x_c$ $\approx$ 0.14 and therefore exhibits the HO phase up to some critical pressure. As with the pure compound URu$_{2}$Si$_2$, the pressure dependence of the feature in $\rho(T)$ increases above some critical pressure $P_c$ near 0.8 GPa. In contrast, for the sample with an Os concentration $x$ = 0.16 (Fig.~\ref{pressure resistivity}(c)) greater than $x_c$, the pressure dependence of the feature in $\rho(T)$ is constant up to 2 GPa suggesting the sample is likely already homogenous in the LMAFM phase at ambient pressure.\\  
\indent The $T_0(P)$ behavior for all seven single crystal samples from the URu$_{2-x}$Os$_x$Si$_2$ system (at $x$ = 0, 0.07, 0.08, 0.15, 0.16, 0.18, and 0.28) is plotted in the ``composite'' $T_0$ vs $P$ phase diagram shown in Fig.~\ref{phase}. The $T_0(P)$ phase boundaries for samples with $x$ = 0, 0.07, and 0.08 exhibit the characteristic discontinuity or ``kink'', which is indicative of the first-order HO$\rightarrow$LMAFM phase transition. The slopes of the $T_0(P)$ phase boundaries in the HO phase, prior to the discontinuities, for the $x$ = 0, 0.07, and 0.08 samples are d$T_0$/d$P$ = 1.11, 0.99, and 1.21 K GPa$^{-1}$, respectively. In the LMAFM phase, the slopes are significantly higher at d$T_0$/d$P$ = 2.99, 2.53, and 2.66 K GPa$^{-1}$, respectively. There is no discontinuity in the slope of the $T_0(P)$ phase boundaries for the Os-substituted samples with higher Os concentrations of $x$ = 0.15, 0.16, 0.18, and 0.28 that are above $x_c$, where the slopes were determined to be d$T_0$/d$P$ = 2.31, 2.42, 2.15 and 2.27 K GPa$^{-1}$, respectively. Note the equivalence between the values of the pressure dependence in both the HO phase (averaged at d$T_0$/d$P$ $\approx$ 1.10 K GPa$^{-1}$) and the LMAFM phase (averaged at d$T_0$/d$P$ = 2.47 K GPa$^{-1}$) across all of the samples. The values of all slopes were determined by linear fits (solid lines in Fig.~\ref{phase}) to the $T_0(P)$ data in the HO or LMAFM phases and are in very good agreement with hydrostatic pressure coefficients reported in other investigations.\cite {Jeffries_2007, Butch_2010a, Kambe_2013, Wolowiec_2016}\\
\begin{figure}[t]
 \includegraphics[width=1.0\linewidth, trim= 2cm 1cm 1.5cm 2cm, clip=true ]{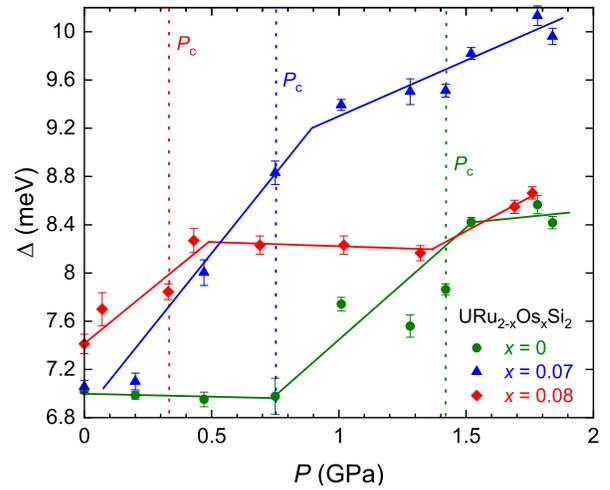} 
\caption{\label{gap}  (Color online) Energy gap $\Delta$ vs.\ pressure $P$ for the $x$ = 0, 0.07, and 0.08 samples. The value of $\Delta$ are based on fits to the low temperature $\rho(T)$ data as explained in the text. The values of $P_{c}$ (marked by dashed vertical lines) were determined from the ``kinks'' in the $T_{0}$ vs.\ $P$ phase boundaries shown in  Fig.~\ref{phase}. The error in $\Delta$ was determined by the fitting algorithm and the solid lines are guides to the eye.} 
\end{figure}
\indent The pressure dependence of the charge gap $\Delta$ that opens up over the Fermi surface during its reconstruction at the PM$\rightarrow$HO/LMAFM phase transition may serve as another measure of the critical pressure $P_{c}$. Namely, the critical pressure $P_{c}$ can be taken as the value of $P$ where there is a change in the pressure dependence of $\Delta$ that occurs at the first-order phase transition from HO to LMAFM. Figure~\ref{gap} displays a plot of the energy gap $\Delta$ as a function of pressure $P$ for single crystal samples of URu$_{2-x}$Os$_{x}$Si$_{2}$ with $x$ = 0, 0.07, 0.08. The values of $\Delta$ were extracted from fits of a theoretical model \cite{Fontes_1999} of electrical resistivity to the $\rho(T)$ data in the low temperature region below the feature in electrical resistivity (where d$\rho$/d$T$ $>$ 0) as described in Ref.~\onlinecite{Wolowiec_2016}. From the $\Delta$ vs.~$P$ plots for each of the $x$ = 0, 0.07, 0.08 samples, there is a flattening of the pressure coefficient d$\Delta$/d$P$ at pressures of $P$ $\approx$ 1.55, 0.90, and 0.50 GPa that are consistent with the critical pressures $P_c$ = 1.43, 0.75, 0.33 GPa determined from the $T_0(P)$ phase boundaries in Fig.~\ref{phase}. Interestingly, the ``kinks'' in $\Delta$ vs.~$P$ data occur at pressures that are consistently $\sim$ 0.15 GPa higher than the $P_c $ values. Low values of $\Delta$ $\approx$ 7.2 meV correspond to transitions into the HO phase. Higher values of $\Delta$ $>$ 8.5 meV correspond to transitions into the LMAFM phase. Intermediate values of $\Delta$ preceding the plateau in the $\Delta$($P$)  curves suggest inhomogeneity and a percolation of the LMAFM phase as pressure is increased.\\  
\begin{figure}[t] 
 \includegraphics[width=1.0\linewidth, trim= 2cm 1cm 2cm 2cm, clip=true]{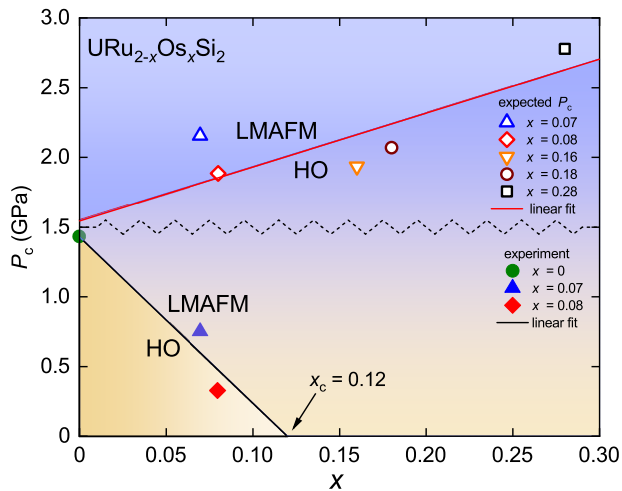}
 \vspace{- 0.4cm}
\caption{\label{critical} (Color online) Measured and expected critical pressure $P_c$ as a function of $x$ for URu$_{2-x}$Os$_{x}$Si$_{2}$. As $x$ is increased, the critical pressure is reduced to $P_c$ = 0 GPa at a critical Os concentration of $x_c$ $\sim$ 0.12. The open symbols represent the expected critical pressure $P_c$ (see text). The black (red) solid lines represent the experimental (expected) HO/LMAFM phase boundaries and are linear fits to the experimental (expected) values of critical pressure $P_c$.} 
\vspace{-.57 cm}
\end{figure} 
\begin{figure*}[t]
   \includegraphics[clip,width=0.36\linewidth]{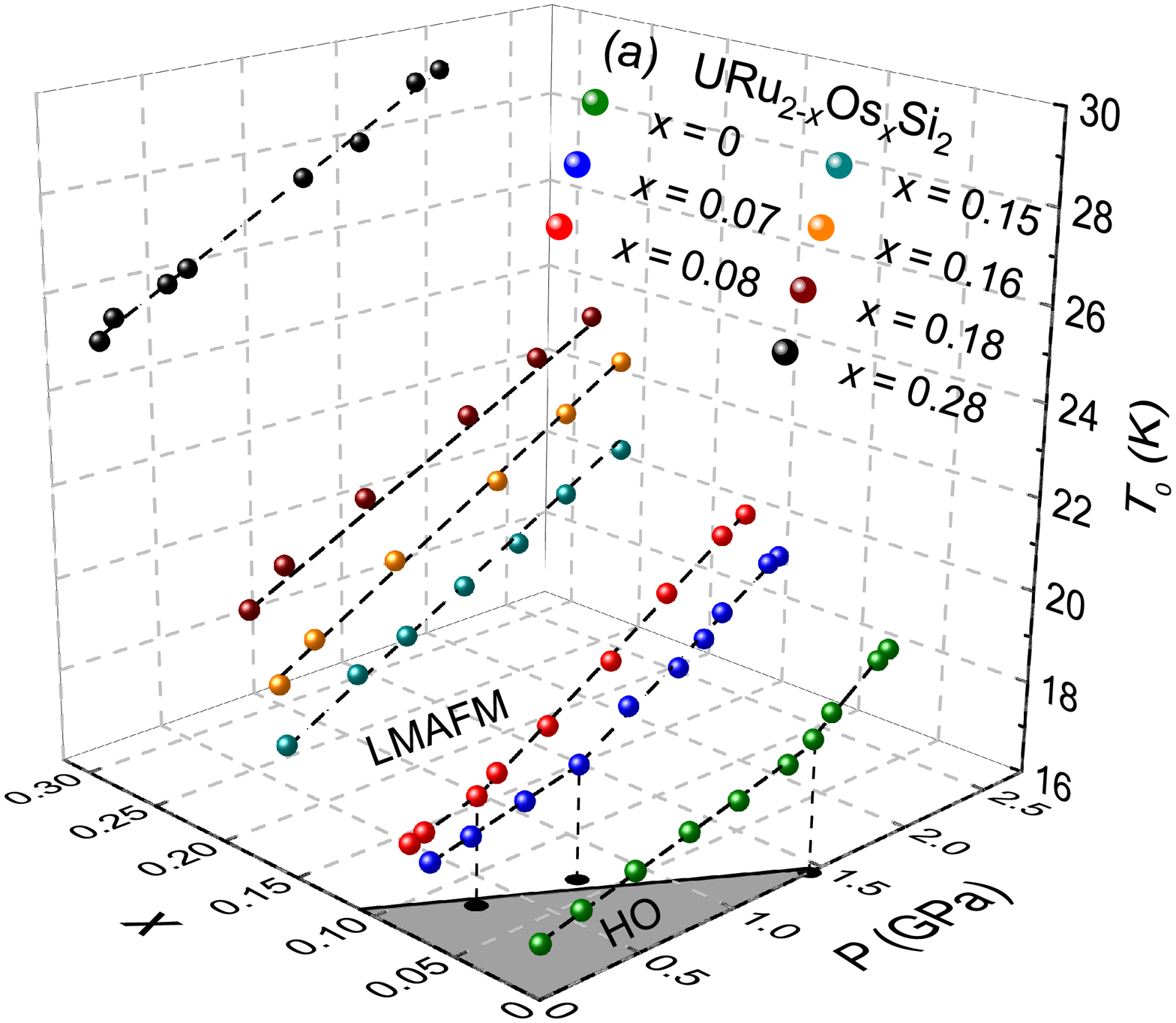}
   \hspace{-0.55cm}
    \includegraphics[clip,width=0.36\linewidth]{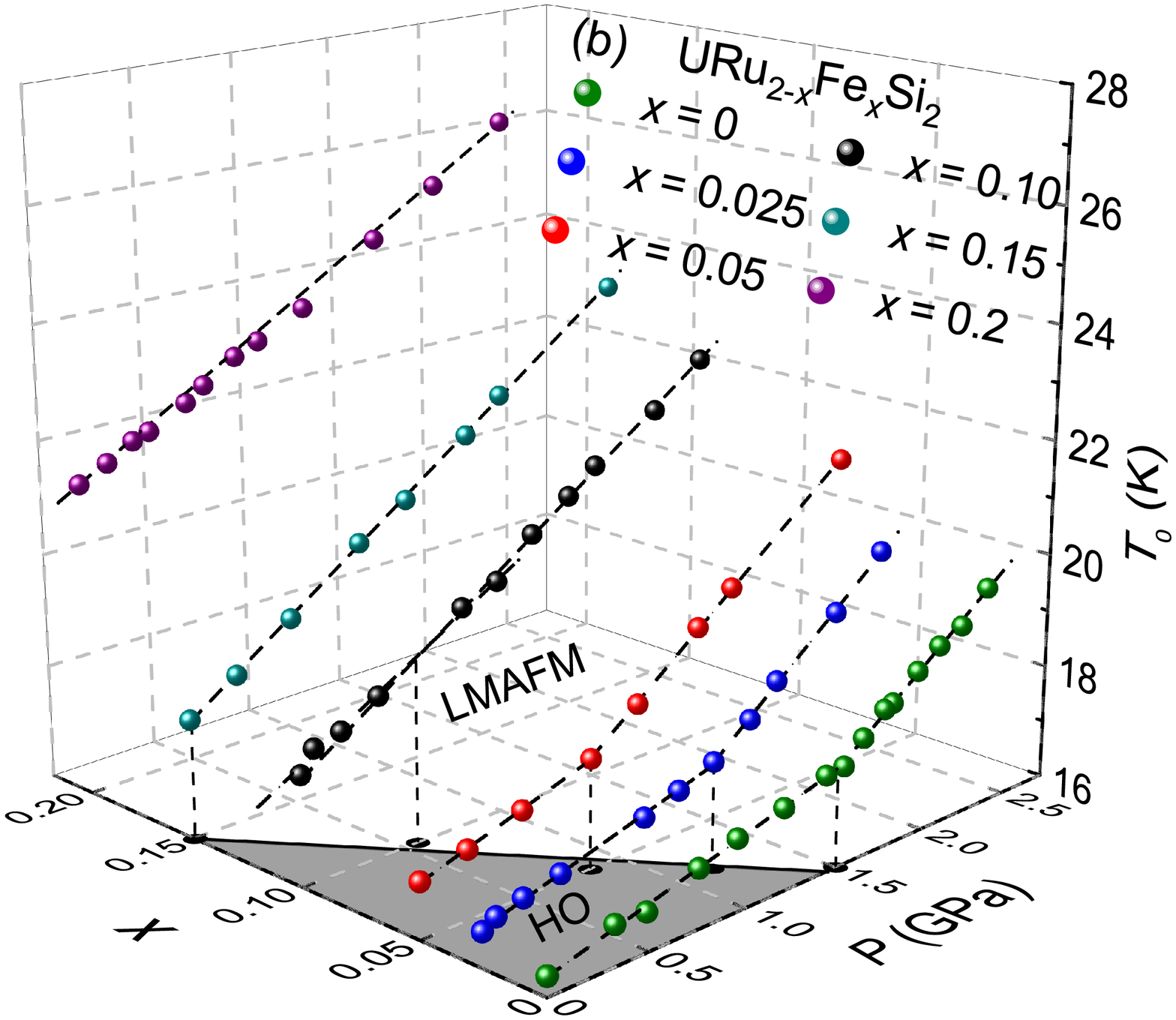}
    \hspace{-1.4cm}
     \includegraphics[clip,width=0.36\linewidth]{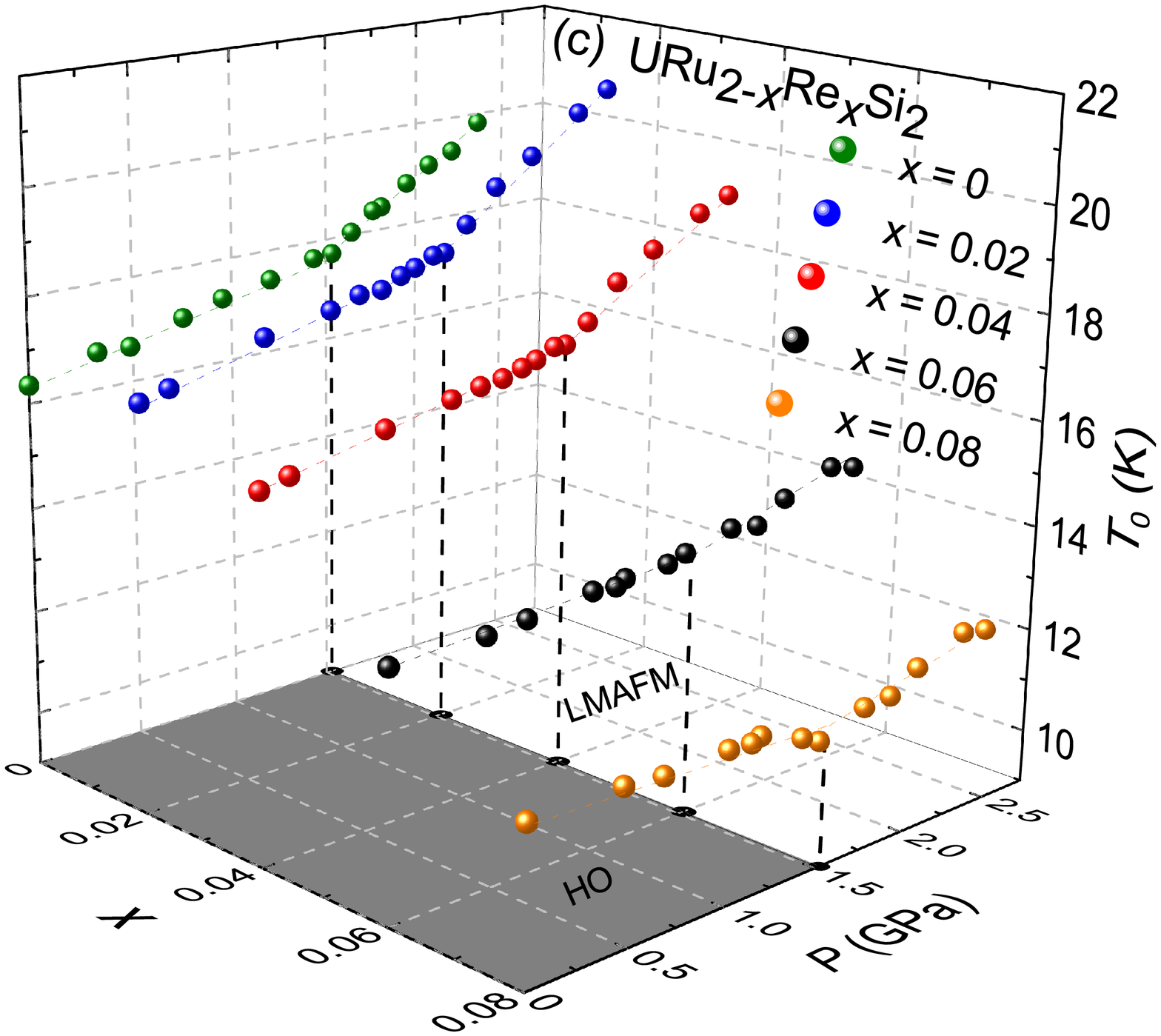}
\caption{\label{3D_phase} (Color online)  $T_{0}$--$P$--$x$ phase diagrams for the (a) URu$_{2-x}$Os$_{x}$Si$_{2}$ (b) URu$_{2-x}$Fe$_{x}$Si$_{2}$ and (c) URu$_{2-x}$Re$_{x}$Si$_{2}$ systems.  The values along the concentration ($x$) axis for the URu$_{2-x}$Re$_{x}$Si$_{2}$ system in panel (c) are reversed relative to those of panels (a) and (b). The $T_0(x,P)$ data for the URu$_{2-x}$Re$_{x}$Si$_{2}$ system were taken from Ref.~\onlinecite{Jeffries_2007}.}  \vspace{-.47 cm}
\end{figure*}
\indent Of central importance to the current report is the reduction of the critical pressure $P_c$ with increasing Os concentration $x$, as illustrated in Fig.~\ref{phase}. The vertical dashed lines locate decreasing values of $P_c$ = 1.43, 0.75, and 0.33 GPa for samples in order of increasing Os concentration $x$ = 0, 0.07, and 0.08. This is reminiscent of the reduction of $P_c$ with increasing Fe concentration for the URu$_{2-x}$Fe$_x$Si$_2$ system.\cite{Wolowiec_2016} Based on the results of the Fe-substituted system, in which lower values of $P_c$ were required to induce the HO$\rightarrow$LMAFM transition according to the additive relation: $P_{ch}(x)$ + $P_c$ $\approx$ 1.5 GPa, one would expect that $higher$ (rather than lower) values of $P_c$ are required to induce the HO$\rightarrow$LMAFM transition for Os-substituted URu$_{2}$Si$_2$, which is biased with an effective $negative$ chemical pressure ($P_{ch}(x)$ $<$ 0). However, this is not what we observed. The discrepancy between the expected increase in $P_c$ and the reduction in $P_c$ that was observed experimentally, is illustrated in the plot of $P_c$ vs.~$x$ as shown in Fig.~\ref{critical}. The solid black line with a negative slope is a linear fit to the experimentally determined values of $P_c$ (filled symbols) and represents the $P_c(x)$  phase boundary between the HO and LMAFM phases for the URu$_{2-x}$Os$_{x}$Si$_{2}$ system. The extrapolation of the fit to zero pressure yields a critical Os concentration of $x$ = 0.12, which is comparable to the value of $x_c$ = 0.14 determined from the ''kink" in $T_{0}$ vs $x$ phase diagram displayed in Fig.~\ref{phase_diagram_x}. The open symbols in Fig.~\ref{critical} represent the expected values of critical pressure $P_c$, which were determined by first converting the Os concentration $x$ to a \textit{negative} chemical pressure $P_{ch}(x)$ and then using the additive property of chemical and applied pressure: $P_{ch}(x)$ + $P_c$ $\approx$ 1.5 GPa. The solid red line with positive slope is a linear fit to these expected values of $P_c$ and represents the expected $P_c(x)$ phase boundary between the HO and LMAFM phases for the URu$_{2-x}$Os$_{x}$Si$_{2}$ system.\\ 
\indent Other than pure URu$_{2}$Si$_2$, Fe-substituted URu$_{2}$Si$_2$, and Os-substituted URu$_{2}$Si$_2$, the only known URu$_{2}$Si$_2$-based system measured under pressure is Re-substituted URu$_{2}$Si$_2$.\cite{Jeffries_2007} At ambient pressure, the effect of Re substitution is to rapidly suppress HO toward an emergent itinerant ferromagnetic phase. Interestingly, as pressure is applied to samples from the URu$_{2-x}$Re$_x$Si$_2$ system, the HO phase is enhanced toward the same HO$\rightarrow$LMAFM phase transition. However, as the Re concentration is increased in URu$_{2-x}$Re$_x$Si$_2$ under pressure, the ``kink'' in the $T_0$ vs.~$P$ ``composite'' phase diagram persists at a critical pressure of $P_c$ = 1.5 GPa. This difference is emphasized in Fig.~\ref{3D_phase} which displays $T_{0}$--$P$--$x$ phase diagrams for each of the URu$_{2-x}$Os$_{x}$Si$_{2}$, URu$_{2-x}$Fe$_{x}$Si$_{2}$ and URu$_{2-x}$Re$_{x}$Si$_{2}$ systems. The $T_0(x,P)$ data for the URu$_{2-x}$Re$_{x}$Si$_{2}$ system was taken from Ref.~\onlinecite{Jeffries_2007}. (Due to the fact that the HO transition temperature $T_0$ is suppressed with increasing Re concentration, the values along the concentration ($x$) axis in Fig.~\ref{3D_phase}(c) have been reversed for clarity.) Note the difference in the HO/LMAFM phase boundary in the  $x$--$P$ plane for the Re-substituted system in Fig.~\ref{3D_phase}(c). The HO/LMAFM phase boundary is constant at $P_c$ = 1.5 GPa for all Re concentrations up to $x$ = 0.08 in URu$_{2-x}$Re$_x$Si$_2$, while the boundary is suppressed to $P$ = 0 GPa as $x$ is increased in the URu$_{2-x}$Fe$_x$Si$_2$ and URu$_{2-x}$Os$_x$Si$_2$ systems.\\ 
\section{Discussion}
\indent Investigations of URu$_{2}$Si$_2$ under applied uniaxial and/or hydrostatic pressure show that an increase in pressure enhances HO (with an increase in $T_0$) and drives the system toward a pressure-induced antiferromagnetic phase (LMAFM) at a critical pressure of $P_c$ $\approx$ 1.5 GPa at the bicritical point (or at $P_x$ $\approx$ 0.5 GPa as $T$ $\rightarrow$ 0).\cite{Amitsuka_1999, Matsuda_2001, Nakashima_2003, Jeffries_2007, Villaume_2008, Butch_2010a, Niklowitz_2010, Bourdarot_2011, Kambe_2013, Williams_2017b} Recently, a related investigation of Fe-substituted URu$_{2}$Si$_2$ under applied pressure established a quantitative equivalence between \textit{positive} chemical pressure $P_{ch}(x)$ to external pressure $P$ in affecting the phase behavior in URu$_{2}$Si$_2$.\cite{Wolowiec_2016} The equivalence between $P_{ch}(x)$ and $P$ is reflected in the consistent  ``additive'' relationship $P_{ch}(x)$ + $P_c$ $\approx$ 1.5 GPa, where the critical pressure $P_c$ necessary to drive the HO$\rightarrow$LMAFM phase transition in URu$_{2-x}$Fe$_{x}$Si$_{2}$ decreases with increasing Fe concentration $x$. The relevance of pressure- and chemical-induced changes to the lattice and how they relate to hybridization between $f$- and $d$-electron states is discussed in more detail below.\\ 
\indent The results presented here for the effect of increasing Os concentration $x$ on the enhancement of HO in URu$_{2-x}$Os$_{x}$Si$_{2}$, as well as the reduction of the critical pressure $P_c$ that induces the HO$\rightarrow$LMAFM transition, are remarkably similar to the phase behavior reported for Fe-substituted URu$_{2}$Si$_2$.\cite{Wolowiec_2016} However, the isoelectronic substitutions of Fe and Os have contrasting effects on the body-centered-tetragonal (bct) lattice. Substitution of smaller Fe ions at the Ru site leads to a contracted lattice and a \textit{positive} chemical pressure $P_{ch}(x)$ in URu$_{2-x}$Fe$_{x}$Si$_{2}$, while substitution of larger Os ions at the Ru site leads to an expanded lattice and a \textit{negative} chemical pressure $P_{ch}(x)$ in URu$_{2-x}$Os$_{x}$Si$_{2}$ (The effect of Os substitution on the lattice is given in Sec.~\ref{appendix}.) This complicates the view of a reduction in the unit-cell volume through applied or chemical pressure as a necessary condition for the enhancement of HO in URu$_{2}$Si$_2$.\\   
\indent Here we suggest an increase in the hybridization of the uranium 5$f$-electron states and transition metal $d$-electron states as the cause for the enhancement of HO toward the HO$\rightarrow$LMAFM phase transition in URu$_{2}$Si$_2$. High-resolution angle-resolved photoemission spectroscopy (ARPES) and scanning-tunneling microscropy (STM) measurements show directly that the HO phase emerges from a paramagnetic (PM) Kondo phase that has clear signatures of hybridization (i.e., Kondo screening) between the localized 5$f$- and itinerant $spd$-electron states, with the onset of hybridization forming at a coherence temperature $T_{coh}$ $\approx$ 70 K.\cite{Santander_2009, Yoshida_2010, Schmidt_2010, Aynajian_2010, Boariu_2013} At lower temperatures close to the HO transition temperature $T_0$, there is an increase in the 5$f$-$d$-electron hybridization leading to a Fermi surface instability as more U-5$f$ electrons dissolve into the FS.\cite{Santander_2009, Yoshida_2010, Schmidt_2010, Aynajian_2010, Boariu_2013, Chatterjee_2013, Meng_2013, Frantzeskakis_2020} The degenerate crossing of hybridized 5$f$-$d$ bands at the Fermi energy $E_F$ create density of states ``hot spots'' or instabilities at the Fermi surface in the PM phase.\cite{Elgazzar_2009, Oppeneer_2010} Hence, small perturbations to the electronic structure in the PM phase may lift the degeneracy and remove the FS instability leading to the opening of a ``hybridization gap'' over roughly 70\% of the FS in the HO and LMAFM phases and a ``re-hybridization'' of the 5$f$- and $d$-electron states. Such a topological reconstruction of the FS is observed during the second-order symmetry-breaking transition (or Lifshitz transition) from the PM phase to the HO (or LMAFM) phase.\\
\indent In this report, we suggest that when URu$_{2}$Si$_2$ is tuned with pressure or with either of the isoelectronic substitutions of Fe or Os at the Ru site, subtle changes occur to the 5$f$-$d$-electron hybridization near the Fermi level which favor the stability of the gapped FS of the HO (or LMAFM) phase over the instability of the FS in the Kondo-like PM phase. As a result, there is an observed increase in the transition temperature $T_0$ with increasing pressure $P$ or substituent concentration $x$. This applies to the observed increase in $T_N$ for the PM$\rightarrow$LMAFM phase transition, during which the FS undergoes a similar reconstruction and gapping. Inelastic neutron scattering experiments performed on single crystals from the URu$_{2-x}$Fe$_x$Si$_2$ system reveal similar interband correlations where enhanced local-itinerant electron hybridization also leads to the stability of the LMAFM phase.\cite{Butch_2016} Below, we address the manner in which each of the three perturbations (pressure, Fe substitution, and Os substitution) independently favors the hybridization of the U-5$f$- and $d$-electron states. Hence, the additivity of $x$ and $P$ in enhancing HO and inducing the LMAFM phase in both the URu$_{2-x}$Fe$_x$Si$_2$ and URu$_{2-x}$Os$_x$Si$_2$ systems is also explained.\\
\indent PRESSURE: Application of uniaxial and hydrostatic pressure both reveal that the pressure dependence of the HO transition temperature $T_0$ is anisotropic with respect to changes in the $a$ and $c$ lattice parameters of the tetragonal crystal. The $a$ lattice parameter (or the shortest U-U separation in the basal plane of the tetragonal lattice) appears to be important in affecting the magnetic properties of URu$_{2}$Si$_2$, as well as the transition to the LMAFM phase.\cite{Bourdarot_2011, Kambe_2013} Furthermore, it has been shown that it is not possible to induce the HO$\rightarrow$LMAFM phase transition with uniaxial stress along the $c$ axis.\cite{Bourdarot_2011} Nor does the ratio of lattice parameters $c$/$a$ appear to be important in governing the salient magnetic properties and phase behavior of URu$_{2}$Si$_2$.\cite{Bourdarot_2011} These pressure-induced changes to the lattice are closely connected to spatial and energetic changes that may occur to the $s$-, $p$-, $d$-, and $f$-electronic orbitals. It is well known that the application of pressure reduces the interatomic distance within a crystal lattice leading to the delocalization and overlapping of electronic orbitals.\cite{Drickamer_1965, Drickamer_1973, Schilling_1979, Schilling_1981} As a consequence, applied pressure can lead to an increase in the hybridization between $f$- and $d$-electron states,\cite{Maple_1969_1, Maple_1970_1, Maple_1976} which is important for the formation of the HO phase, and is now considered to be one of its defining characteristics.\cite{Santander_2009, Chatterjee_2013, Meng_2013} Here, we suggest that the pressure-induced enhancement of hybridization in URu$_2$Si$_2$-based systems contributes to the instability at the FS that leads to the gapping of the FS and the second-order transition to the HO and LMAMF phases.\\  
\indent Fe-SUBSTITUTION: The remarkable agreement between $P_{ch}(x)$ and $P$, and their effect on the HO and LMAFM phases is not a surprise, considering that Fe substitution results in an almost entirely uniaxial contraction along the $a$ parameter axis. Upon substitution of smaller Fe ions for Ru, it is suggested that the effective chemical pressure $P_{ch}$ associated with the reduction in the interatomic spacing, favors increased overlap and hybridization of the U-5$f$-electron states and $d$-electron states in much the same way that applied pressure $P$ favors hybridization.\cite{Kanchanavatee_2011, Wolowiec_2016} Hence, the sum result for investigations of URu$_{2}$Si$_2$ under pressure and investigations of Fe substitution in URu$_{2-x}$Fe$_{x}$Si$_{2}$ suggest that a contracted lattice in the direction of the $a$ axis is necessary for the enhancement of HO and a transition to LMAFM in URu$_{2}$Si$_2$.\cite{Kanchanavatee_2011, Janoschek_2015, Wolowiec_2016}. In addition to the comparable effects of Fe substitution and pressure on the lattice, HO, and LMAFM, we discuss below the binding energy of the Fe-4$d$ electrons as a relevant factor for the increase in 5$f$- and $d$-electron hybridization.\\ 
\indent Os-SUBSTITUTION: In contrast, the effective ``negative'' chemical pressure associated with an expanded crystal lattice upon substitution of larger Os ions for Ru should not favor hybridization of the U-5$f$- and $d$-electron states in URu$_{2-x}$Os$_x$Si$_2$. However, an increase in the hybridization may still occur if one considers one or both of the following: (1) The larger spatial extent of the 5$d$-electron orbitals in osmium compared to that of the 4$d$-electron orbitals in ruthenium where the change in the radius of the $d$-electron wave functions is 0.639 \AA~for Ru ions to  0.706 \AA~for Os ions.\cite{Mann_1973, McLean_1981} In the tight-binding approximation, the overlap for a pair of orbitals is dominated by an exponential term which decays on a length scale given by the inverse sum of the radii of the two electronic orbitals.\cite{Papaconstantopoulos_2003, Durgavich_2016, Papaconstantopoulos_2015} The increase in the radius of the $d$-electron wave functions when an Os ion replaces a Ru ion is significant and would considerably effect the overlap of the $d$-electron wave functions and U-5$f$-electron wave functions. (2) The stronger spin-orbit coupling that occurs in Os compared to that of Ru may lead to a broadening of the $d$-electron energy bands and an increase in the number of $d$ electrons at the Fermi level.\cite{Oppeneer_2010}\\
\vspace{-.2cm}
\subsection*{Calculations of  $f$- and $d$-electron hybridization in U$M_2$Si$_2$ with $M$ = (Fe, Ru, and Os)}
\begin{figure}[t] 
 \includegraphics[width=1.0\linewidth, trim= 2cm 1cm 2cm 2cm, clip=true]{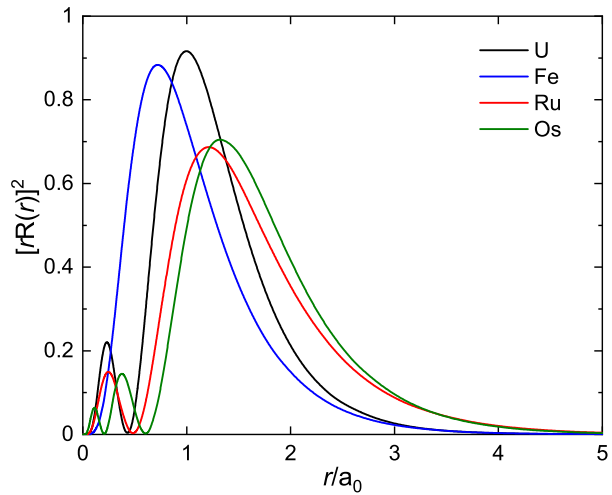}
 \vspace{- 0.4cm}
\caption{\label{Fig7} (Color online) The radial probability distribution for the $f$- and $d$-electrons in U, Fe, Ru, and Os ions, expressed in atomic units.} 
\vspace{-.57 cm}
\end{figure} 
\indent In an effort to further understand the hybridization between the U 5$f$-electron states and the transition metal $d$-electron states in the U$M_2$Si$_2$ series with $M$ = (Fe, Ru, and Os), we performed tight-binding calculations of the overlap of the U-5$f$-electron states and $d$-electron states of the Fe, Ru, and Os ions. 
Preliminary calculations of Hartree-Fock wave-functions were made for the U-5$f$-electron states, the partially-filled $d$-electron shells of the transition metal ions Fe,  Ru, and Os, as well as the 3$p$-electron states of Si and 4$p$-electron states of Ge. The radial probability distributions for the $f$- and $d$-electron states are displayed in Fig.~\ref{Fig7}. Relative to the 4$d$-electron states of Ru, the 3$d$-electron states of Fe are more localized, whereas the 5$d$-electron states of Os are slightly more spatially extended.\\ 
\indent Additionally, the Hartree-Fock binding energies $E_{HF}$ for the $f$- and $d$-electron orbitals in U, Fe, Ru, and Os were calculated and are summarized in Table~\ref{binding table}. Note that there is a non-monotonic trend in the binding energy for the transition metal ions (Fe, Ru, and Os) moving down the column in the periodic table with $E_{HF}$ = 1.2157, 0.80244, and 1.0095 $ry$ for the Fe 3$d^6$, Ru 4$d^7$, and Os 5$d^6$ states, respectively. The anomalous electronic configuration of atomic Ru sets its binding energy much lower than that of the $d$-electrons of Fe, Os. The binding energy of Ru is also much lower than that of the $f$-electrons of U where $E_{HF}$ = 1.2689 $ry$ for U 5$f^3$ state.\\
\begin{table}[t]
\caption{Hartree-Fock binding energies $E_{HF}$ of  $f$- and $d$-electron orbitals in U, Fe, Ru, and Os ions, where the energies are given in Rydbergs (Ry).}          
\centering
 \noindent\rule{8.7cm}{1pt}\\       
     \begin{tabular}{ c  c}\\
     Element~~~~~~~~&~~~~~~~~Binding energy $E_{HF}$ (Ry)\\ [2ex]
     \hline\\
    U 5$f^3$~~~~~~~&~~~~~~~~1.2689\\[2ex]
    \hline\\
    Fe 3$d^6$~~~~~~~&~~~~~~~~1.2157\\ 
    Ru 4$d^7$~~~~~~~&~~~~~~~~~0.8024\\
    Os 5$d^6$~~~~~~~&~~~~~~~~1.0096\\ [2ex]
    \end{tabular}
    \noindent\rule{8.7cm}{1pt}
    \label{binding table}
    \end{table} 
\indent Tight-binding energy matrix elements $t_{\alpha,\beta}(R)$ were calculated to determine the degree of overlap of the U-5$f$-electron states with the Fe, Ru, and Os $d$-electron states for atomic sites separated by a distance $R$. 
Hybridization energies $\Delta_{\alpha}(R)$ were estimated from the matrix elements $t_{\alpha,\beta}(R)$ and from the calculated Hartree-Fock binding energies for the U, Fe, Ru, and Os ions in URu$_{2-x}$$M_x$Si$_2$ with $M$ = (Fe, Os) according to the expression: 
\begin{equation}
\label{hybridization equation}
\Delta_{\alpha}(R) = \sum_{\beta}\frac{|t_{\beta,\alpha}(R)|^2}{E_{5f} - E_{d,\beta}},
\end{equation}
where $E_{5f}$ = 1.2689 $ry$ is the binding energy for the U 5$f^3$ state and $E_{d,\beta}$ = 1.2157, 0.8024, and 1.0095 $ry$ are the binding energies for the Fe 3$d^6$, Ru 4$d^7$, and Os 5$d^6$ states, respectively (see Table~\ref{binding table}).\\  
\indent Table~\ref{hybridization table} contains the hybridization energies as a measure of the degree of hybridization between the U 5$f$-electron states and the $d$-electron states of Fe, Ru, and Os. The hybridization energies $\Delta_{\alpha}(R)$ are smallest for the hybridization of Ru $d$-electrons, which suggests a diminished hybridization for the $d$ electrons of the Ru ions compared to those of the Fe and Os ions. Hybridization of the Os $d$-electron states with the U 5$f$-electron states is largest, being only slightly larger than that of the Fe $d$-electron states. This ordering of $\Delta_{\alpha}(R)$ for Fe, Ru, and Os is attributed to both the increasing spatial extent of the $d$-electron wave function down the column of the periodic table (Fig.~\ref{Fig7}) and also the non-monotonic variation in excitation energy (or binding energy). However, the non-monotonic variation in binding energy is the dominant effect, where the binding energy of the Ru $d$-electrons is much lower than that of the Fe and Os $d$-electrons and also the $f$-electrons of U (see Table~\ref{binding table}).\\ 
\indent Similar trends in 4$f$-$d$-electron hybridization are reported for the heavy-Fermion and Kondo-like systems of  CeFe$_2$Si$_2$, CeRu$_2$Si$_2$, and CeRu$_{2-x}$Os$_x$Si$_2$, where the strength of the hybridization of the Ce-4$f$ electrons and the $s$, $p$, and $d$ conduction electrons can be characterized by the Kondo temperature $T_K$.\cite{Flouquet_1988} CeFe$_2$Si$_2$ has a large Kondo temperature $T_K$ $\sim$ 103 K,\cite{Koterlyn_2007} while the Kondo temperature for CeRu$_2$Si$_2$ is $T_K$ $\sim$ 10 to 25 K.\cite{Besnus_1985, Umarji_1986, Kitaoka_1986, Flouquet_1988} As small amounts of Os are introduced into CeRu$_{2-x}$Os$_x$Si$_2$, the Kondo temperature increases to $T_K$ $\sim$ 10$^2$ K  for $x$ = 0.1.\cite{Umarji_1986, Godart_1986} These changes in the hybridization of the Ce-4$f$ and $s$, $p$, and $d$ electrons across the CeFe$_2$Si$_2$, CeRu$_2$Si$_2$, and CeRu$_{2-x}$Os$_x$Si$_2$ systems appear to be consistent with the changes in the 5$f$-$d$-electron hybridization in other reports \cite{Amorese_2020} and with our calculations across the U$M_2$Si$_2$ series with $M$ = (Fe, Ru, and Os).\\
 
\begin{table}[t]
\caption{Hybridization energies of the $\alpha$-th 5$f$-orbital with the $\beta$th $d$-orbital, where the energies are given in Rydbergs (Ry).}          
\centering
 \noindent\rule{8.7cm}{1pt}\\       
     \begin{tabular}{ c  c  c  c}\\
    $\Delta_{\alpha}(R)$ &~~~~~~~~Fe (Ry) &~~~~~~~~Ru (Ry) &~~~~~~~~~~Os (Ry)\\ [2ex]
     \hline\\
    $\Delta_{xyz}(R)$  &~~~~~~~~0.081&~~~~~~~~0.034&~~~~~~~~~~0.097\\ 
    $\Delta_{x(5x^2-3r^2)}(R)$  &~~~~~~~~0.058&~~~~~~~~0.025&~~~~~~~~~~0.064\\ 
    $\Delta_{y(5y^2-3r^2)}(R)$   &~~~~~~~~0.050&~~~~~~~~0.021&~~~~~~~~~~0.055\\
    $\Delta_{z(5z^2-3r^2)}(R)$   &~~~~~~~~0.079&~~~~~~~~0.033&~~~~~~~~~~0.087\\ 
    $\Delta_{x(y^2-z^2)}(R)$   &~~~~~~~~0.039&~~~~~~~~0.018&~~~~~~~~~~0.047\\ 
    $\Delta_{y(z^2-x^2)}(R)$   &~~~~~~~~0.039&~~~~~~~~0.018&~~~~~~~~~~0.047\\
    $\Delta_{z(x^2-y^2)}(R)$   &~~~~~~~~0.067&~~~~~~~~0.028&~~~~~~~~~~0.084\\ [2ex]
    \end{tabular}
    \noindent\rule{8.7cm}{1pt}
    \label{hybridization table}
    \end{table}
Hence, the enhancement of the HO phase in URu$_{2-x}$Os$_x$Si$_2$ with increasing Os concentration $x$ is consistent with the greater degree of $d$- and $f$-electron hybridization as calculated for the Os ions. Similar reasoning may also explain the enhancement of HO in the case of URu$_{2}$Si$_2$ under applied pressure $P$ and the case of URu$_{2-x}$Fe$_x$Si$_2$ with increasing Fe concentration $x$. The reduction of the critical pressure $P_c$, and  the cooperative effects of $x$ and $P$ observed in URu$_{2-x}$Os$_x$Si$_2$, may follow from the nature in which both the perturbations of $x$ and $P$ work together to foster hybridization:  applied pressure favors delocalization of the U-5$f$ electrons and the substitution of Os ions for Ru extends the $d$ electrons outward within the unit cell. Both of these effects together would favor overlap between the U-5$f$- and $d$-electron wave functions in URu$_{2-x}$Os$_x$Si$_2$.\\
\indent The increase in spin-orbit coupling may also help with hybridization of the U-5$f$- and Os-5$d$-electron states on account of the splitting of the $d$-electron band, which brings the orbitals closer together in energy and slightly enhances the hybridization between the two orbital levels with $j$ = $l - 1$, where $l = 3$ for U and $l = 2$ for Os. The increase in hybridization between the U-5$f$-electron states and the transition metal $d$-electron states, caused by the larger spin-orbit coupling of Os, is estimated to be limited and less than $\sim$ 2 \%.\cite{Papaconstantopoulos_2015}\\ 
\indent The persistence of the critical pressure at $P_c$ = 1.5 GPa, with increasing rhenium (Re) concentration in URu$_{2-x}$Re$_x$Si$_2$, suggests that any doping which suppresses HO may not be ``additive'' with pressure, and, as such, is not a perturbation that favors hybridization. Indeed, for small Re concentration ($x$ $<$ 0.1) in URu$_{2-x}$Re$_x$Si$_2$, the hidden order transition $T_0$ is rapidly reduced and for higher Re concentrations ($x$ $>$ 0.1), the system enters a ferromagnetic state rather than the LMAFM phase.\\
\indent Based on our hybridization calculations and previous reports of the trends in 5$f$-$d$-electron hybridization for the  3$d$-, 4$d$-, and 5$d$-electron orbitals, one might expect the same qualitative increase in hybridization (relative to the Ru4$d$ electrons) for the Re-5$d$-electron states as observed for the Os-5$d$-electron states. However, the trends in 5$f$-$d$-electron hybridization reported here for U$M_{2}$Si$_2$ ($M$ = Fe, Ru, and Os) and elsewhere for Ce$M_{2}$Si$_2$ ($M$ = Fe, Ru, and Os)\cite{Besnus_1985, Umarji_1986, Kitaoka_1986, Godart_1986, Flouquet_1988} are for systems that are isoelectronic. For these systems, there is little or no variation across the series in the number of $d$-band electrons near the Fermi energy that are available for hybridization. The degree of $f$-$d$-electron hybridization is largely dependent on the density of states at the Fermi level such that any significant variation in the number of $d$ electrons near $E_F$ would have an effect on the hybridization. \cite{Dalichaouch_1990, Amorese_2020}  Furthermore, substitutions for Ru such as Rh and Re that are effectively electron (or hole) doping would shift the Fermi energy away from the degenerate crossing of the hybridized bands thereby stabilizing the FS in the paramagnetic phase. In addition, any doping resulting from non-isoelectronic substitutions might also change the underlying band structure and shape of the FS, which experimentally is shown to disrupt HO and replace it with an un-ordered state.\cite{Oh_2007, Balicas_2007}\\
\indent Hence, there are competing effects on hybridization in moving from Ru to Re, where any increase in hybridization owing to the spatially extended character of Re-5$d$ electrons is mitigated by the reduction in the number of $d$ electrons available near $E_F$ for hybridization and other deleterious effects to the FS. The $d$ orbital electronic configuration is 5$d^5$ for Re compared to 5$d^6$ for Os. In the case of Re substitution, the decrease in the density of $d$-band electrons (or increase in hole concentration) may inhibit the 5$f$-$d$-electron hybridization that is observed to increase in the case of Os substitution. In addition, the degree of hybridization between U-5$f$ electrons and Re-5$d$ electrons depends not only on the hybridization matrix elements but also largely on the binding energy of the Re-5$d$ electrons (see Equation~\ref{hybridization equation}). Hence, a comparison of hybridization across systems that are not isoelectronic is more complicated and it is not unreasonable to assume that hybridization in the case of Re substitution would not increase as observed for the case of Os substitution. For systems in which the HO phase is suppressed with increasing substituent $x$, as in Re-substituted URu$_{2}$Si$_2$, a determination of the hybridization between the U-5$f$ and Re-5$d$ electrons as a function of concentration $x$ should be investigated further. 
\section{Conclusions}
\indent Early specific heat measurements of URu$_2$Si$_2$ in 1985 revealed an anomalous feature at $T_0$ = 17.5 K, reminiscent of a continuous mean-field type of phase transition.\cite{Maple_1986}  The use of a simple model for the analysis of the specific heat anomaly led to the notion of a partial gapping of the Fermi surface as the compound entered the hidden order (HO) phase, with the magnitude of the gap determined to be 11 meV.\cite{Maple_1986} This simple yet powerful experimental technique was one of the first ``probes'' into the structure or ``reconstruction'' of the Fermi surface during the HO phase transition in URu$_2$Si$_2$. Over the last 20 years, advanced experimental techniques have yielded direct evidence and provided confirmation of the partial gapping of the Fermi surface, with gap values of $\sim$ 10 meV. We now have a detailed picture of the electronic structure in proximity to the hidden order transition at $T_0$, whereby the onset of hybridization of 5$f$ and $d$ electrons at 70 K leads to a degenerate crossing of 5$f$-$d$-hybridized bands at the Fermi level and ultimately to an instability, reconstruction, and partial gapping of the Fermi surface at 17.5 K.\\ 
\indent Currently, applied pressure, and the substitution of Fe and Os ions for Ru, are the only known perturbations to URu$_2$Si$_2$ that result in an enhancement of HO and a subsequent first-order transition to the LMAFM phase. Here, we explain the enhancement of HO as the result of an increase in the hybridization of the uranium 5$f$- and transition metal (Fe, Ru, Os) $d$-electron states, which leads to a Fermi surface instability that favors the HO phase over the PM phase. This causes the increase in the  PM$\rightarrow$HO transition temperature $T_0$.\\
 \indent The results from transport measurements for single crystals of URu$_{2-x}$Os$_x$Si$_2$ under pressure presented here are used to construct the $T_0$($x$, $P$) phase behavior. As the concentration of Os is increased, there is both an observed increase in $T_0$ and a reduction in the critical pressure $P_c$ necessary to induce the transition to the LMAFM phase. This is consistent with previously reported effects of applied pressure and Fe substitution on HO and $P_c$ in single crystals of URu$_{2-x}$Fe$_x$Si$_2$. However, the expansion of the lattice with increasing Os concentration is anomalous in comparison to the lattice contraction observed in URu$_2$Si$_2$ with increasing pressure and in URu$_{2-x}$Fe$_x$Si$_2$ with increasing Fe substitution.\\ 
\indent Hence, the increase in the 5$f$- and $d$-electron hybridization appears to be dependent on various effects, both spatial and energetic. The contraction of the lattice with pressure or chemical pressure tends to favor both the overlap and hybridization of electronic orbitals, whereas the spatially extended $d$-electron orbitals (as with Os 5$d$ electrons) can also lead to an increase in their hybridization with the ``localized'' 5$f$ electrons. In this report, results of tight-binding calculations show that the degree of hybridization between the U 5$f$ electrons with the transition metal $d$ electrons is largely dependent on the difference in binding energy between the ``localized '' 5$f$ electrons and $d$-band electrons. In general, it is noted that the trend in hybridization increases in moving away from the Ru 4$d$ electrons to the Fe 3$d$ and Os 5$d$ electrons. This is true for other isoelectronic systems such as Ce$M_{2}$Si$_2$ ($M$ = Fe, Ru, and Os).     

\section{Appendix: Sample Quality}
\label{appendix}
There are some differences among the single crystal samples in the temperature dependence of the electrical resistivity at ambient pressure at the PM$\rightarrow$HO/LMAFM transition (see Fig.~\ref{ambient resistivity}). The features at the transition temperature $T_0$ are relatively sharp for the $x$ = 0, 0.07, 0.08, and 0.16 samples, whereas for $x$ = 0.15, 0.18, and 0.28 samples, the transition at $T_0$ is broadened. These differences in the $\rho(T)$ behavior warrant some clarification and additional comments regarding the quality of the single crystal samples. We note that there will be some unavoidable and intrinsic broadening of the transition for larger concentrations of osmium. This type of broadening is also seen in the polycrystal data for Os concentrations larger than $x$ = 0.2 previously reported in Ref.~\onlinecite{Kanchanavatee_2014} and also in other work reported in Ref.~\onlinecite{Hall_2015}, in which the feature in the $\rho(T)$ data at $T_0$ is nearly non-existent for the $x$ = 0.1 sample. The broadening of the transition at $T_0$ may be attributed to inhomogeneity that arises when single crystals are grown out of melts in a tetra-arc furnace by the Czochralski method. As larger concentrations of Os are introduced into the melt, there is a larger chance of the occurrence of inhomogeneity across the sample.\
\begin{figure}[t] 
\includegraphics[width=1.0\linewidth, trim= 1.3cm 1cm 0.5cm 1cm, clip=true]{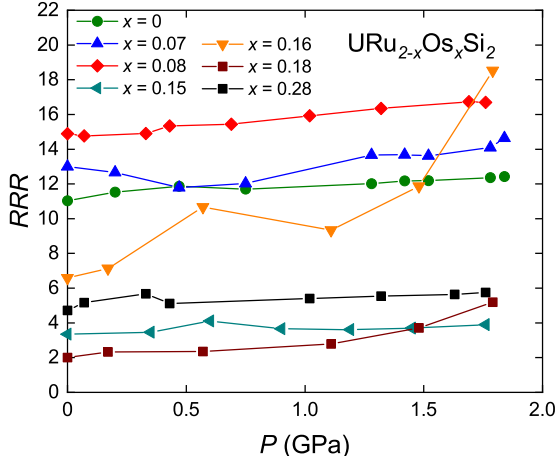}
\caption{\label{rrr} (Color online)  Values of the residual resistivity ratio (RRR) defined as $\rho(300 K)/\rho(2 K)$ as a function of pressure $P$ for single crystal samples with $x$ = 0, 0.07, 0.08, 0.15, 0.16, 0.18, 0.28.}
\vspace{- 0.6 cm}
\end{figure} 

\begin{figure}[b]
\includegraphics[scale = 0.20, trim= 2cm 3.4cm 1cm 2cm, clip=true]{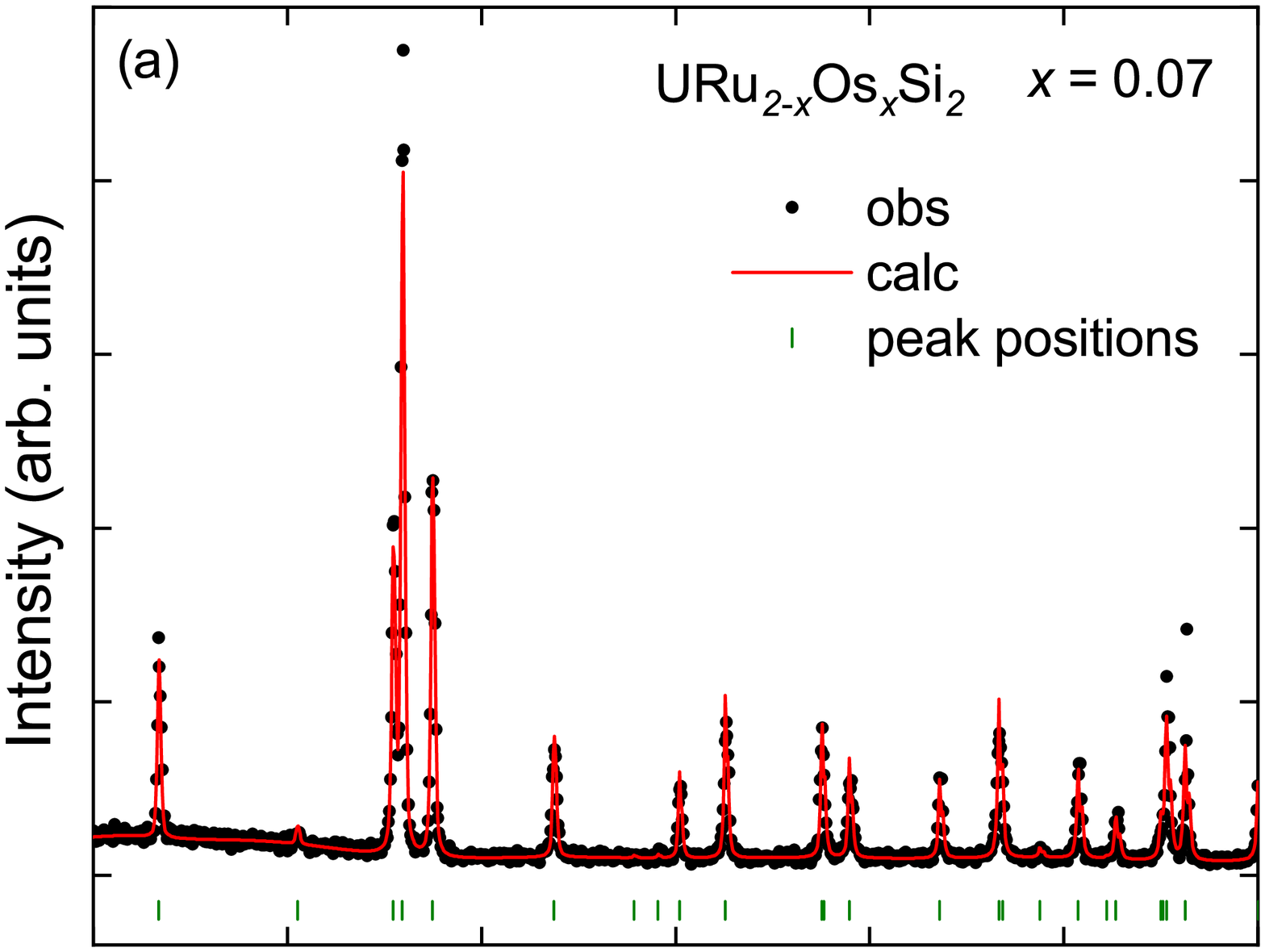}
\vspace{-0.1 cm}
\includegraphics[scale = 0.20, trim= 2cm 3.4cm 1cm 2.4cm, clip=true]{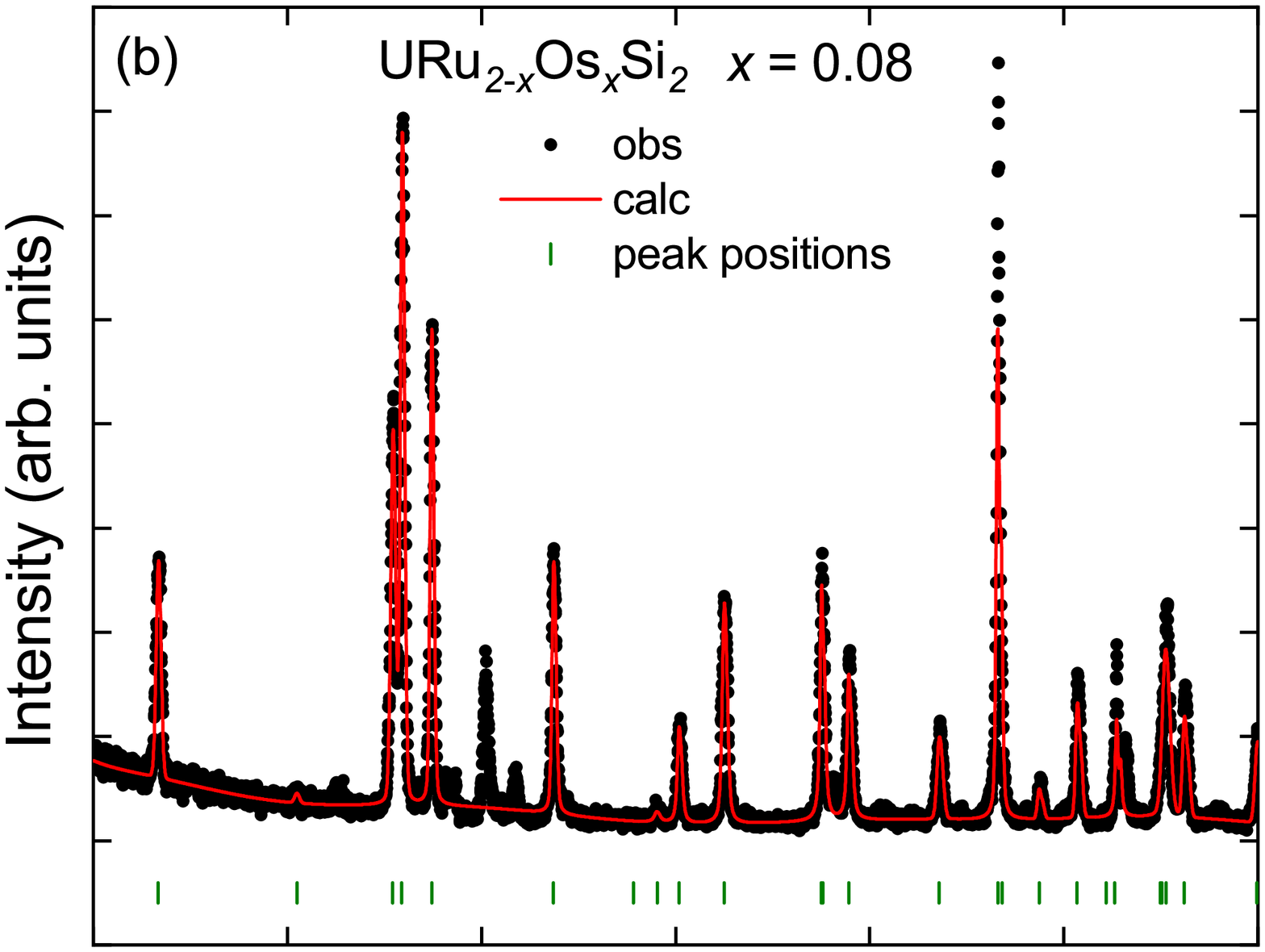}
\vspace{-0.1 cm}
\includegraphics[scale = 0.20, trim= 2cm 3.4cm 1cm 2cm, clip=true]{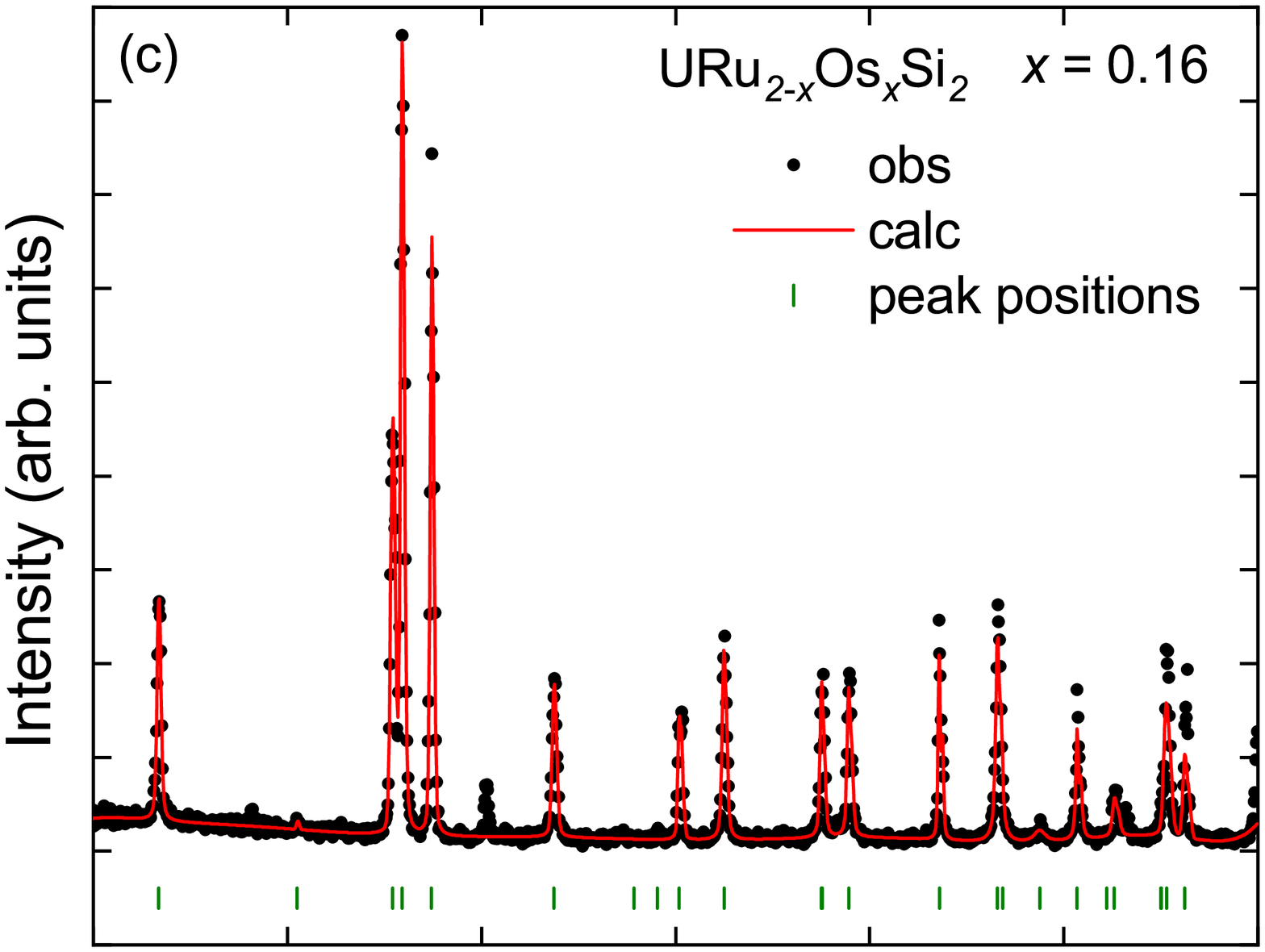}
\vspace{-0.1 cm}
\includegraphics[scale = 0.20, trim= 2cm 3.4cm 1cm 2cm, clip=true]{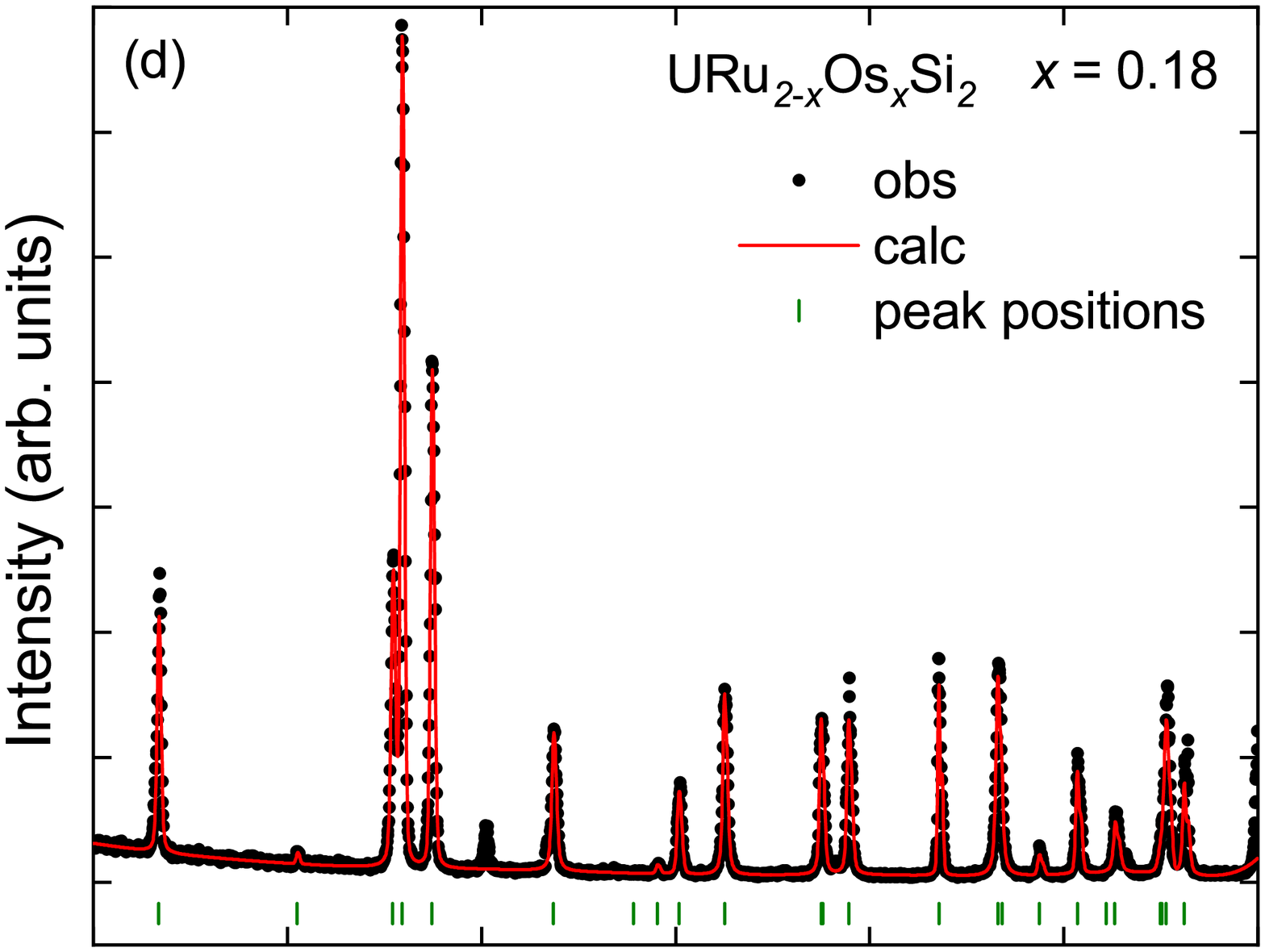}
\includegraphics[scale = 0.20, trim= 2cm 1cm 1cm 2cm, clip=true]{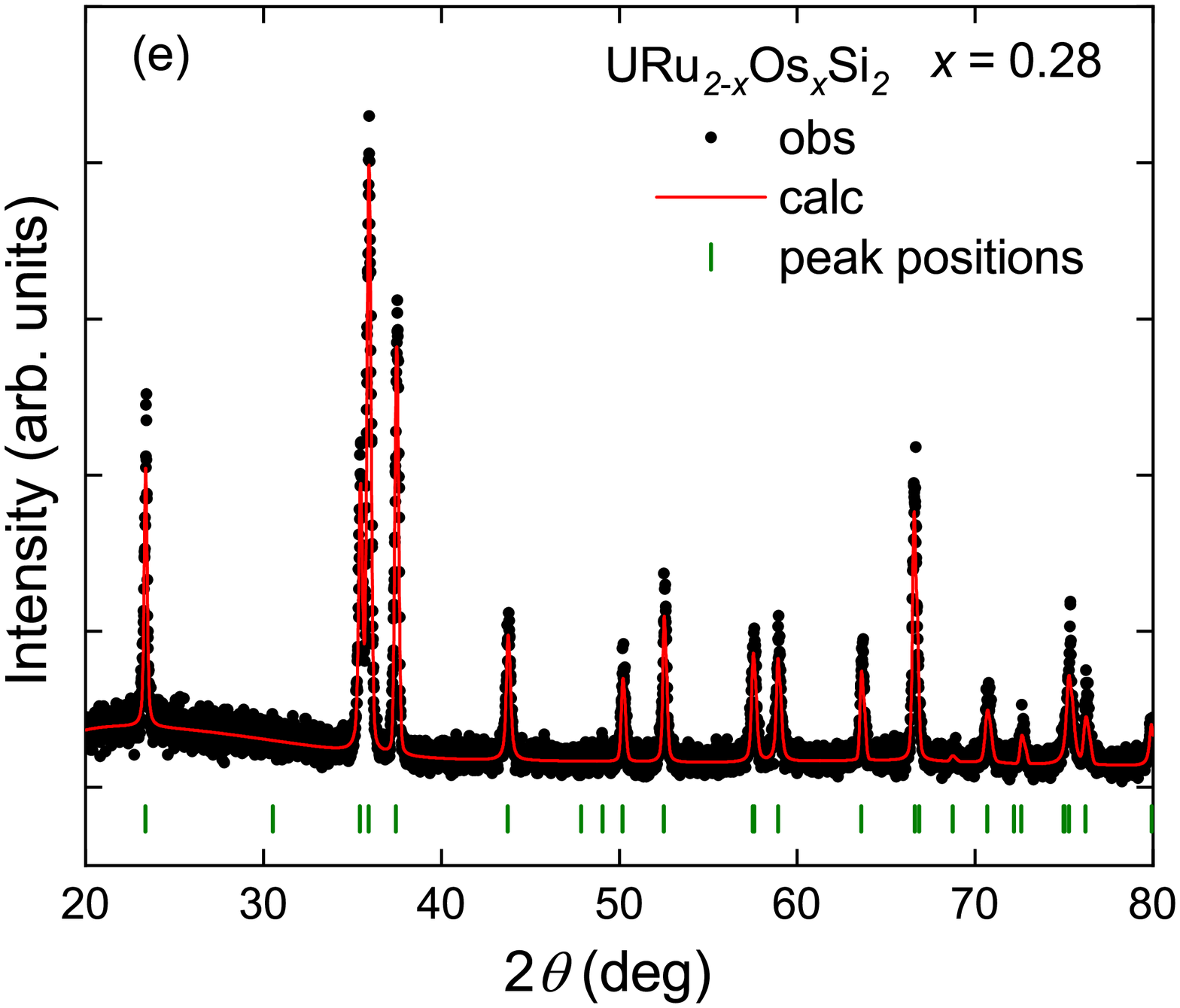}
\vspace{-0.1 cm}
\includegraphics[scale = 0.21, trim= 1.1cm 1cm 0.6cm 2cm, clip=true]{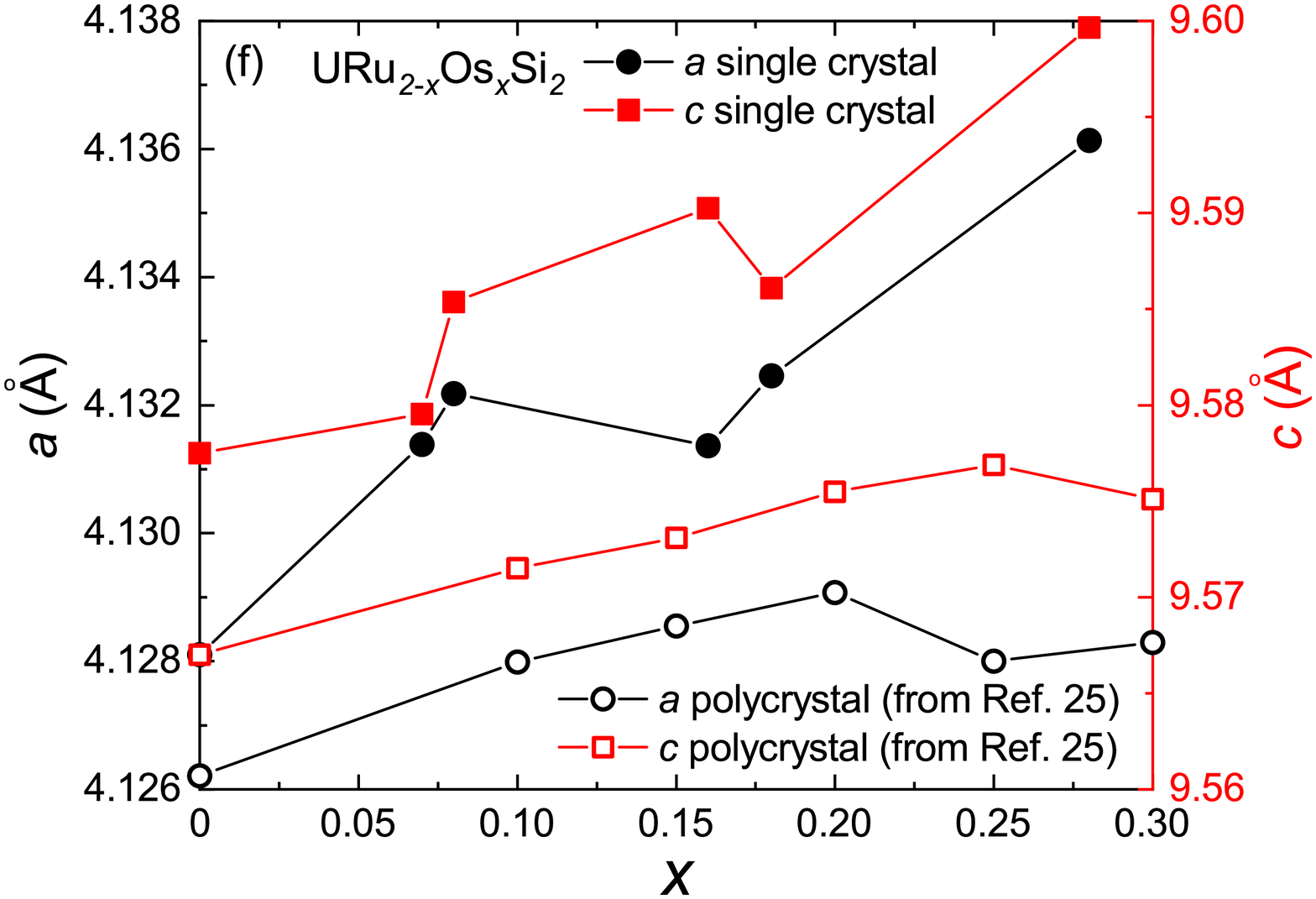}
\vspace{-0.2 cm}
\caption{\label{refinements}  (Color online) (a)--(e) X-ray diffraction patterns for single crystal samples of URu$_{2-x}$Os$_x$Si$_2$ with Os concentrations $x$ = 0.07, 0.08, 0.16, 0.18, 0.28. The red line is the Rietveld refinement fit to the data in black. (f) Lattice parameters $a$ and $c$ versus osmium concentration $x$.} 
\vspace{- 0.5cm}
\end{figure}

\indent The broadening of the transition at $T_0$ also occurs with increasing pressure. From the $\rho(T)$ data displayed in Fig.~\ref{pressure resistivity}(a), the degree of broadening appears to be monotonic with increasing pressure. This type of pressure-induced broadening of the PM$\rightarrow$HO transition has been previously reported for polycrystalline URu$_{2}$Si$_2$ (see Fig. 1 in Ref.~\onlinecite{McElfresh_1987}), where the broadening of the $\rho(T)$ feature at $T_0$ may be associated with ``different states of strain within the polycrystalline sample''.\cite{McElfresh_1987} Similar broadening of the HO transition occurred in single crystals of URu$_{2}$Si$_2$  ($x$ = 0) under applied hydrostatic pressure (see Ref. \onlinecite{Jeffries_2007}). Broadening of the transition at $T_0$ is also seen in single crystal samples of URu$_{2-x}$Fe$_x$Si$_2$ as a function of increasing Fe concentration $x$ (see Ref.~\onlinecite{Wolowiec_2016}). Hence, the broadening of the $T_0$ transition observed in the single crystal samples of URu$_{2-x}$Os$_x$Si$_2$ with $x$ = 0.18 and 0.28 (see Fig.~\ref{ambient resistivity}) is not to be unexpected. However, the broadening in the $x$ = 0.15 sample is somewhat anomalous and needs some clarification.\\ 
\begin{figure}[t] 
\includegraphics[width=0.7\linewidth, trim= 2.5cm 2cm 3.7cm 1cm, clip=true]{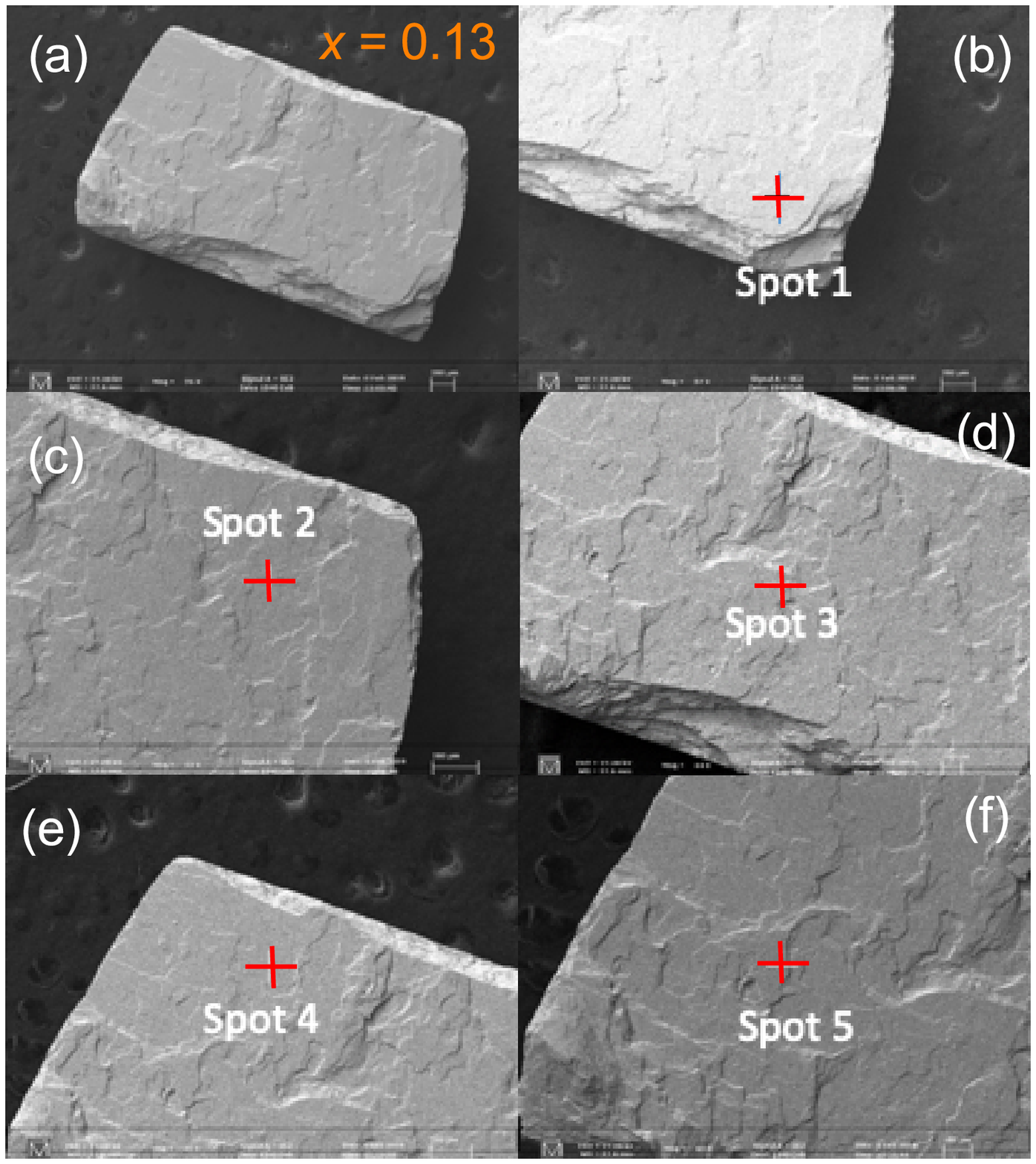}
\vspace{-0.2 cm}
\includegraphics[width=0.8\linewidth, trim= 2cm 1cm 0.3cm 2cm, clip=true]{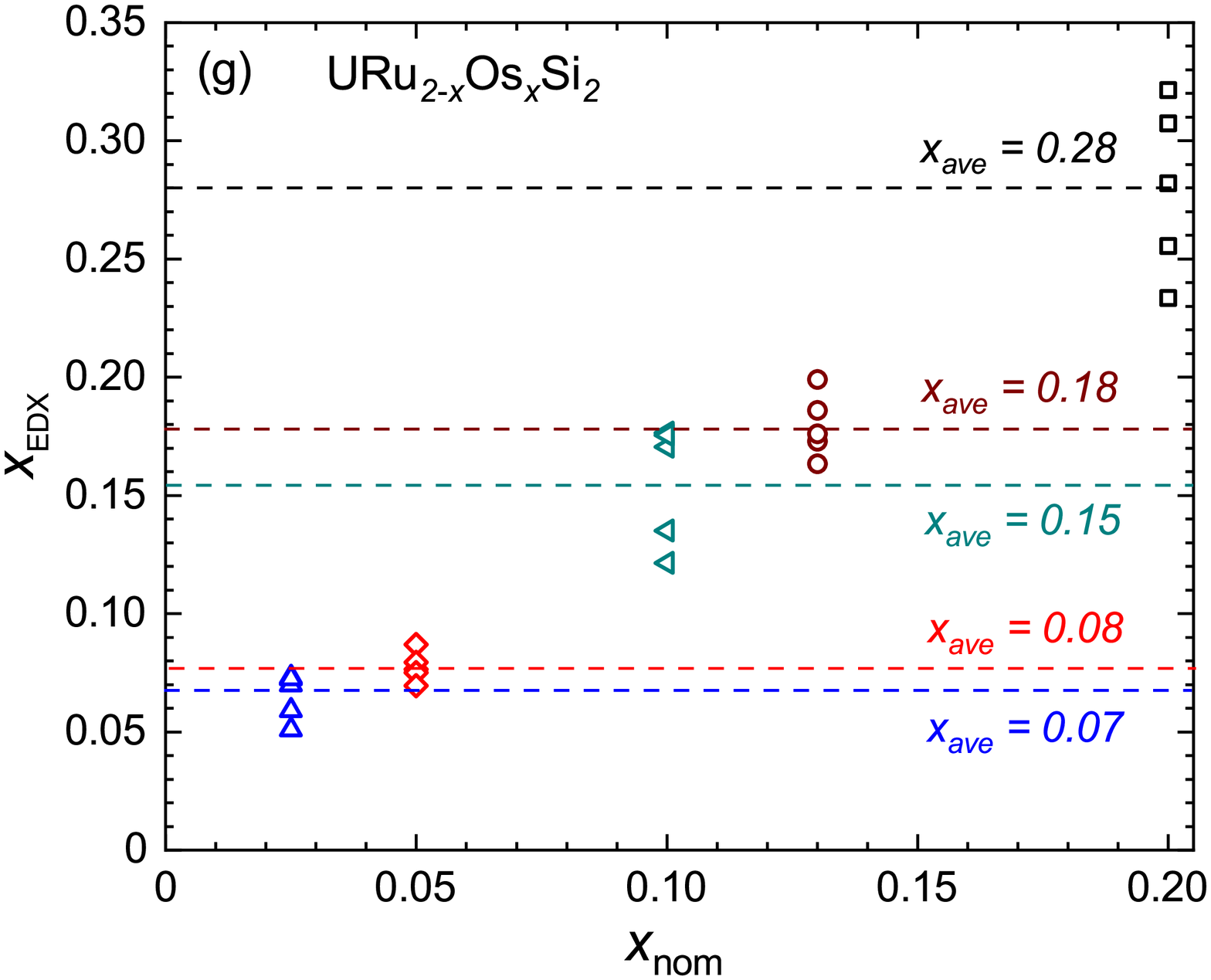}
\caption{\label{EDX} (Color online)  Results of EDX measurements for single crystal samples of URu$_{2-x}$Os$_x$Si$_2$ with nominal concentrations of $x_{nom}$ = 0.025, 0.05, 0.10, 0.13, and 0.20. (a) Image of the single crystal sample with $x_{nom}$ = 0.13. (b)--(f) An illustrative example of a set EDX measurements at five different spots over the surface of the $x_{nom}$ = 0.13 sample. (g) Plot of the five osmium concentrations $x_{EDX}$ from EDX measurements versus $x_{nom}$ for each sample. Horizontal dashed lines are located at the average Os concentration $x_{ave}$ of the five $x_{EDX}$ values and are taken as the actual Os concentration $x_{act}$ = $x$ for each sample.}
\end{figure}    
\indent As a measure of sample quality, we determined the residual resistivity ratio (RRR), defined as $\rho(300 K)/\rho(2 K)$, for each of the single crystal samples. Values of RRR as a function of pressure $P$ are displayed in Fig.~\ref{rrr} for single crystal samples with Os concentration $x$ = 0, 0.07, 0.08, 0.15, 0.16, 0.18, 0.28. Note that for all samples, RRR increases with $P$, which is likely due to the improvement in the contact resistance between the leads and the sample as pressure is increased. At ambient pressure, the values of RRR are larger than 10 for those samples with lower Os concentration ($x$ = 0, 0.07, 0.08). In comparison, significantly lower values of RRR = 3.3, 2, and 4.7 for the $x$ = 0.15, 0.18, and 0.28 samples suggest more inhomogeneity or the presence of impurity-induced electron scattering, which is likely to cause some broadening of the $\rho(T)$ feature at the transition temperature $T_0$.\\
\indent X-ray diffraction (XRD) patterns for single crystal samples of URu$_{2-x}$Os$_x$Si$_2$ with Os concentrations $x$ = 0.07, 0.08, 0.16, 0.18, 0.28 are shown in Fig.~\ref{refinements}(a)--(e). The black circles represent the data, the red line represents the fit results from Rietveld refinement to the data, and the green ticks indicate peak positions for the URu$_{2}$Si$_2$ crystal. The peaks in the data at 2$\theta$ = 40$^{\circ}$ for $x$ = 0.08 in panel (b), $x$ = 0.16 in panel (c), and $x$ = 0.18 in panel (d) are Os impurities. The results from the fits of the Rietveld refinement to the XRD data were used to determine the lattice constants $a$ and $c$ for single crystal samples of URu$_{2-x}$Os$_x$Si$_2$ with $x$ = 0.07, 0.08, 0.16, 0.18, 0.28. The lattice constants $a$ and $c$ for single crystals (filled symbols) as well as for polycrystalline samples (open symbols) are plotted in Fig.~\ref{refinements}(f) as a function of osmium concentration up to $x$ = 0.3. (The polycrystal data was taken from Ref.~\onlinecite{Kanchanavatee_2014}.) For the single crystal samples, both lattice constants $a$ and $c$ are shown to increase by approximately 0.2\% as Os concentration $x$ is increased in URu$_{2-x}$Os$_x$Si$_2$. For the single crystal samples, there is some degree of scatter in the expansion of both $a$ and $c$ between $x$ = 0.08 and 0.18, which may be related to the presence of Os impurity peaks detected in the XRD patterns for the $x$ = 0.08, 0.16, and 0.18 samples (see Fig.~\ref{refinements}).\\
\indent Single crystal samples of URu$_{2-x}$Os$_x$Si$_2$ at nominal concentrations of $x_{nom}$ = 0.025, 0.05, 0.10, 0.13, and 0.20 were measured for elemental composition using energy-dispersive X-ray spectroscopy (EDX). For each of these samples, EDX measurements were made at five different spots across the surface of the sample. As a representative example, an image of a sample from the $x$ = 0.13 single crystal is shown in Fig.~\ref{EDX}(a) and the locations of the five different EDX measurement spots for this sample are shown in the images displayed in Fig.~\ref{EDX}(b)--(f). The results of the EDX measurements for these samples are shown in Fig.~\ref{EDX}(g), where the Os concentrations ($x_{EDX}$) determined from the five different EDX measurements on each sample are plotted (open symbols) versus the nominal Os concentration ($x_{nom}$). The horizontal dashed lines represent the average Os concentration $x_{ave}$ of the five different EDX measurements for each sample, where $x_{ave}$ is taken to be the actual osmium concentration $x_{act}$ for these single crystal samples. Based on the EDX measurements, the single crystal samples of URu$_{2-x}$Os$_x$Si$_2$ with nominal concentrations of $x_{nom}$ = 0.025, 0.05, 0.10, 0.13, and 0.20 were determined to have actual concentrations of  $x_{act}$ = 0.07, 0.08, 0.15, 0.18, and 0.28, respectively. The error in the Os concentration for each sample is taken as the standard deviation of the five $x_{EDX}$ values and is represented by the error bars in the $T_0$--$x$ phase diagram shown in Fig.~\ref{phase_diagram_x}. Note that for samples with larger concentrations of Os $x$, there is an increase in the error of $x$ suggesting that the inhomogeneity across the sample increases as $x$ increases. For the sample that was unavailable for EDX measurement, with nominal Os concentration of $x$ = 0.16, the error in the Os concentration was taken as the average of the error for the samples with comparable Os concentration, namely $x$ = 0.15, 0.18, and 0.28.

\section*{Acknowledgements}
Research at the University of California, San Diego was supported by the US Department of Energy (DOE), Office of Basic Energy Sciences, Division of Materials Sciences and Engineering, under Grant DE-FG02-04ER46105 (materials synthesis and characterization), the US National Science Foundation (NSF) under Grant DMR 1810310 (low-temperature measurements), and the National Nuclear Security Administration under the Stewardship Science Academic Alliance Program through the US DOE under Grant DE-NA0002909 (high-pressure measurements). Research at the National High Magnetic Field Laboratory (NHMFL) was supported by NSF Cooperative Agreement DMR-1157490, the State of Florida, and the DOE. 
\bibliography{References_Osmium}
\end{document}